\documentclass[a4paper]{jpconf}
\usepackage[pdftex]{graphicx}
\usepackage{wrapfig}
\usepackage{float}
\usepackage{caption}
\usepackage{subcaption}
\usepackage{multirow}
\usepackage[hidelinks]{hyperref}
\usepackage[dvipsnames]{xcolor}
\usepackage{amsmath}
\usepackage{tikz}

\begin{document}
\pagenumbering{arabic}
\title{Implementation of time-dependent Hartree-Fock in real space}

\author{Uday Panta}

\address{Department of Physics, University of California, Merced, 5200 N. Lake Rd., Merced, CA 95343, USA}

\author{David A. Strubbe}

\address{Department of Physics, University of California, Merced, 5200 N. Lake Rd., Merced, CA 95343, USA}

\ead{dstrubbe@ucmerced.edu}
\thispagestyle{plain}
\pagestyle{plain}

\begin{abstract}
Time-dependent Hartree-Fock (TDHF) is one of the fundamental post-Hartree-Fock (HF) methods to describe excited states. In its Tamm-Dancoff form, equivalent to Configuration Interaction Singles, it is still widely used and particularly applicable to big molecules where more accurate methods may be unfeasibly expensive. However, it is rarely implemented in real space, mostly because of the expensive nature of the exact-exchange potential in real space. Compared to widely used Gaussian-type orbitals (GTO) basis sets, real space often offers easier implementation of equations and more systematic convergence of Rydberg states, as well as favorable scaling, effective domain parallelization, flexible boundary conditions, and ability to treat model systems. We implemented TDHF in the Octopus real-space code as a step toward linear-response hybrid time-dependent density-functional theory (TDDFT), other post-HF methods, and ensemble density-functional theory methods involving exact exchange. Calculation of HF's non-local exact exchange is very expensive in real space. We overcome this limitation with Octopus' implementation of Adaptively Compressed Exchange (ACE), and find the appropriate mixing scheme and starting point to complete the ground-state calculation in a practical amount of time, and thus enable TDHF. We compared our results to those from GTOs on a set of small molecules and confirmed close agreement of results, though with larger deviations than in the case of semi-local TDDFT. We find that convergence of TDHF demands a finer real-space grid than semi-local TDDFT. We also present the subtleties in benchmarking a real-space calculation against GTOs, relating to Rydberg and vacuum states.
\end{abstract}

\section{Introduction}
Hartree-Fock (HF) theory \cite{lykos1963discussion, Szabo} is a main pillar in quantum chemistry, serving as the starting point for calculating the electronic structure of molecules.
HF calculations provide molecular geometry, electron distribution, and ground-state properties \cite{doi:https://doi.org/10.1002/9780470125823.ch2}. Once the ground state has been calculated, excited states can be calculated using the ground-state results. A first approximation one can make for the excitation energies is the eigenvalue differences of the unoccupied and occupied orbitals, termed the Independent Particle Approximation (IPA) \cite{yu2016perspective}. Configuration Interaction (CI), Equation-of-Motion Coupled Cluster (EOM-CC), and Time-Dependent Hartree-Fock (TDHF) are examples of post-Hartree-Fock  methods \cite{loos2018mountaineering} that can provide more accurate excited states. 
The exact-exchange potential used in HF is orbital-dependent while it is density-dependent in traditional semi-local Kohn-Sham Density Functional Theory (DFT). Hybrid functionals which are a mixture of these two types of exchange potentials are widely used \cite{becke1993new,perdew1996rationale} to improve accuracy, so TDHF is a building block for time-dependent density-functional theory (TDDFT) with hybrids \cite{dreuw2005single}.

Hartree-Fock theory provides a reasonably accurate description of the ground state of molecular systems, especially for small-to-moderately sized closed-shell molecules and systems where electron correlation effects are not dominant \cite{dreuw2005single}. 
Gaussian-type orbitals (GTOs) are typically used as basis sets for Hartree-Fock \cite{hehre1972self}. Implementation of Hartree-Fock in real space is rare, though implementations on special grids for atoms and diatomic molecules \cite{KOBUS2013799} and a finite-element-type implementation have been published  \cite{kim2015configuration}. Using GTOs conveniently allows integrals to be done analytically. However, the nature of orbitals one can obtain in electronic structure calculations depends on the type and size of the basis set used. The basis set is typically optimized for the ground state which can make description of unoccupied orbitals difficult \cite{dunning1977gaussian}.
This issue is particularly acute for Rydberg states which are highly excited states that are far from the atoms and weakly bound \cite{reisler2009interacting}, but transitions from valence to Rydberg states can be  important in optical properties \cite{paterson2020rydberg}. To include these states, additional diffuse functions are commonly added to the basis sets \cite{kendall1992electron}. Comparisons of GTOs and real space have shown that such very large, diffuse basis sets can be needed to describe the excited states involved in non-linear optical response \cite{Vila}. In real space, Rydberg states do not require special diffuse functions but can be captured by systematically converging the physical size of the real-space domain in which the orbitals are calculated. This domain size and the spacing of the grid-points are the parameters controlling the accuracy of a real-space calculation, which can be systematically converged \cite{kronik2006parsec}. Other advantages of real-space calculations include favorable scaling for large systems, effective parallelization over domains, and straightforward use of both finite and periodic boundary conditions \cite{kronik2006parsec,andrade2012time,andrade2015real}, as well as the ability to handle non-atomic model systems \cite{HFC11}.
In real-space however, the calculation of exchange integrals becomes very expensive \cite{gygi} and these integrals are a major part of any HF calculation underlying TDHF.

In this paper, we present results from our real-space implementation of TDHF in the Octopus code. In section \ref{Theory}, we discuss the general theory behind ground-state HF and TDHF. Section \ref{Exchange_pot} discusses the exact-exchange potential which is the main feature of HF that sets it apart from Kohn-Sham DFT. As this is the most expensive part of the calculation, we discuss the Adatively Compressed Exchange (ACE) formalism to speed up the exact-exchange calculation. We also present the choice of parameters used in our calculations in sections \ref{PPs} and \ref{globalPara}. In section \ref{results} we discuss the general difficulties in reaching a solution to the HF ground state in real space and how to overcome them. Finally, section \ref{Benchmarking} shows the results of TDHF calculations, with comparison to excitation energies obtained from GTOs, demonstrating a good level of agreement and practical attainability of convergence with respect to the real-space grid parameters.

\section{Theory}\label{Theory}
Ground-state HF calculations involve solving for single-particle energy eigenvalues $\{\epsilon_i\}$ and corresponding eigenstates $\{\psi_i\}$ with a self-consistent field (SCF) calculation via the Hartree-Fock-Roothan equations \cite{casida2012progress}:
\begin{equation}\label{roothan}
    \left(-\frac{1}{2}\nabla^2 + V_{\rm{ion}} + V_{\rm H}[\rho] +\hat{V}_{\rm X}[\{\psi_j\}]\right)\psi_i = \epsilon_i\psi_i
\end{equation}
The four terms in the parentheses are the kinetic energy operator, external potential, Hartree potential, and exchange potential. The Hartree potential depends on the density $\rho \left( \textbf{r} \right) = \sum_j^{N_e} \left| \psi_j \left( \textbf{r} \right) \right|^2$, where $N_e$ is the number of electrons. The exchange potential is used in its exact form and its action on an arbitrary orbital $\psi_p$ is given by
\begin{align}\label{exchange_op}
    \hat{V}_{\rm X}[\{\psi_j\}] \psi_p(\textbf{r})  &= -\sum_{j=1}^{N_e} \psi_j(\textbf{r})\int\frac{\psi_j(\textbf{r}')\psi_p(\textbf{r}')}{\vert \textbf{r} - \textbf{r}'\vert}d\textbf{r}'.
\end{align}
Correlation is absent, unlike DFT which has the exchange-correlation potential $V_{\textrm{XC}}$.
To calculate the excited-state properties of a system in most linear-response methods, we need extra unoccupied (virtual) orbitals. Self-consistency is not required in generating these unoccupied orbitals because they do not contribute to the density or exchange potential of the system. %

TDHF can be applied in a real-time propagation form like in TDDFT \cite{Yabana}, but we focus here on its linear-response matrix form. It has a similar structure to linear-response TDDFT,
as an equation for excitation-energy eigenvalues $\omega$ and eigenvectors with excitation component $X$ and de-excitation component $Y$, and block matrices $A$ and $B$ \cite{dreuw2005single}:
\begin{equation}\label{main_eqn}
\begin{bmatrix}\mathbf{A} & \mathbf{B} \\ \mathbf{B^*} & \mathbf{A^*} \\ \end{bmatrix} \begin{bmatrix} \mathbf{X} \\ \mathbf{Y} \\ \end{bmatrix} = \omega \begin{bmatrix} \mathbf{1} & \mathbf{0} \\ \mathbf{0} & -\mathbf{1} \\ \end{bmatrix} \begin{bmatrix} \mathbf{X} \\ \mathbf{Y} \\ \end{bmatrix} .
\end{equation}
The matrix elements in the case of spin-polarized calculations, based on unrestricted HF (UHF) calculations, are given by
\begin{align}\label{casidaA}
    A_{ia,jb} &= \delta_{ij}\delta_{ab}(\epsilon_a - \epsilon_i) + (ia\vert\vert jb)
\end{align}
and 
\begin{align}\label{casidaB}
    B_{ia,jb} &= (ia\vert\vert bj)
\end{align}
where $i$, $j$ represent occupied orbitals and $a$, $b$ represent unoccupied orbitals.
Here the shorthand notation $(pq\vert \vert rs) = (pq\vert rs) -(pr\vert qs)$ has been used, where
\begin{align}
(pq\vert rs) &= \int \frac{d^3\textbf{r}'d^3\textbf{r}}{\lvert\textbf{r}-\textbf{r}^{'}\rvert} \phi_p^*(\textbf{r}') \phi_q(\textbf{r}') \phi_r^*(\textbf{r}) \phi_s(\textbf{r}). %
\end{align}
For spin-unpolarized calculations, based on restricted HF (RHF) ground-state calculations, the matrix elements are given by \cite{casida2012progress}
\begin{align}\label{RHF_matrix}
    A_{ia,jb} &= \delta_{ij}\delta_{ab}(\epsilon_a - \epsilon_i) + \delta_M (ai\vert jb) - (ab\vert ji)
\end{align}
where $\delta_M = 2$ and $0$ for singlets and triplets respectively. The exchange integral is $\left(ai|jb\right)$ and the Coulomb integral is $\left(ab|ji\right)$. The factor $\delta_M$ comes from the cancellation or addition of the exchange term in triplets $\frac{1}{\sqrt2} \left( | \uparrow \downarrow > + | \downarrow \uparrow> \right)$ or singlets $\frac{1}{\sqrt2} \left( | \uparrow \downarrow > - | \downarrow \uparrow > \right)$, respectively \cite{mayer2013simple, RevModPhys.36.844}, which also occurs in TDDFT \cite{casida2012progress, andrade2015real} and the Bethe-Salpeter equation \cite{CohenLouie}.
 
This full TDHF formulation in equation \ref{main_eqn} is a non-Hermitian eigenvalue problem, with matrix size twice the number of transitions considered. For real orbitals, the equation can be transformed into a Hermitian problem of half the matrix size \cite{casida1995time}.
One widely used approximation (applicable for complex orbitals too) is neglecting the matrix $B$ in equation \ref{main_eqn} completely and only solving $AX = \omega X$, which is again a Hermitian problem of half the matrix size. This Tamm-Dancoff approximation (TDA) applied to TDHF is equivalent to Configuration Interaction Singles (CIS) \cite{dreuw2005single}. TDA  not only makes the calculation less computationally intensive, but also can perform better than full TDHF \cite{chantzis2013tamm} in many cases.
Neglecting the exchange and Hartree integrals in equation \ref{RHF_matrix} and calculating the excitation energy as the difference of virtual and occupied eigenvalues is the IPA. This work will focus on TDA-TDHF, adapting Octopus' existing implementation \cite{andrade2015real} of 
HF and Tamm-Dancoff TDDFT \cite{Hirata}.

Once the TDHF eigenvalue problem is solved, one can obtain the absorption spectrum too. Using the TDA singlet excitation eigenvectors $\{x_{k,ia}\}$, the many-body dipole transition matrix elements reduce to a linear combination of single-particle matrix elements \cite{casida2012progress}:
\begin{align}
    \textbf{d}_{k} = \sum_{ia} \left<i\vert \hat{\textbf{r}} \vert a\right> x_{k,ia}
\end{align}
like in TDDFT, where the indices $ia$ run over the pairs of occupied and unoccupied orbitals.
The isotropically averaged oscillator strength, in the electric dipole approximation, is then calculated as
\begin{align}\label{osc_strength}
    f_{k} = \frac{2m_e}{3\hbar^2}\omega_{k} \left| \textbf{d}_{k} \right|^2
\end{align}
The absorption spectrum for unpolarized light is represented by the strength function:
\begin{align}\label{strength_function}
    S(\omega)=\sum_{k} f_{k} \delta(\omega - \omega_{k})
\end{align}
The $\delta$-function is in practice broadened for plotting with a Lorentzian or Gaussian function.

\section{Implementation}\label{Implementation}

\subsection{Exchange Potential and Adaptively Compressed Exchange (ACE)}\label{Exchange_pot}
Unlike semi-local exchange and correlation density functionals such as the Local Density Approximation (LDA)  \cite{PhysRevB.45.13244,dirac1930note},
the Hartree-Fock exchange potential is dependent on the molecular orbitals (equation \ref{exchange_op}) instead of the density.
The non-local nature of this integral makes it time-consuming to solve numerically, especially in methods with large basis sets. Therefore real-space and plane-wave implementations \cite{Castro} use iterative algorithms which only require the application of $\hat{V}_{\rm X}$ to specific orbitals instead of the full construction of the operator $\hat{V}_X$.

Equation \ref{exchange_op} can be solved as a Poisson equation with
$\rho(\textbf{r})=\psi_j(\textbf{r})\psi_i(\textbf{r})$, using various optimized Poisson solvers such as interpolating scaling functions, like for the Hartree potential. The efficiency of these solvers, including their parallelization, has been studied in detail in the context of Octopus in Ref. \cite{garcia2014survey}.
The cost of solving equation \ref{exchange_op} by direct integration for each $r$ is $\mathcal{O}(N_\mu^2)$, while the cost of solving as a Poisson problem is $\mathcal{O}(N_\mu \log N_\mu)$, where $N_\mu$ is the number of grid points.
This is a significant improvement, but exact exchange still remains considerably more computationally intensive than a semi-local functional.

We will use the Adaptively Compressed Exchange (ACE) \cite{lin2016adaptively} approach, which was introduced to speed up the calculation of exact exchange. This method was previously implemented in Octopus. ACE is closely related to the density-fitting technique commonly used with GTOs in quantum chemistry \cite{Werner}.
Other possible approaches to speed up  the exchange calculation include  the Selected Column Density Matrix \cite{scdm} and Recursive Subspace Bisection \cite{gygi} methods. In the ACE formalism, the full-rank exchange operator $\hat{V}_{\rm X}$ is approximated by the low-rank ACE operator, $\hat{V}_{\rm X}^{\rm ACE} = \zeta \zeta^{\rm T}$, which is obtained iteratively as follows.

For ground-state SCF, we start with a guess of the exchange operator calculated using equation \ref{exchange_op}:
\begin{equation}\label{ACE1}
    W_i(\textbf{r}) = (\hat{V}_\textrm{X}[\{\phi\}]\phi_i)(\textbf{r})
\end{equation}
for all $\{\phi_i\}_{i=1}^{N_e}$.
Then matrix $M$ is calculated using $M_{kl} = \int \phi_k(\textbf{r}) W_l(\textbf{r})d\textbf{r}$. Cholesky factorization is performed on $-M$ to get $L$ such that $M=-LL^{\rm T}$, where $L$ is a lower triangular matrix. The projection vector in the ACE formulation is then calculated as
\begin{align}\label{ACE2}
    \zeta_k(\textbf{r}) = \sum_i^{N_e} W_i(\textbf{r})(L^{-{\rm T}})_{ik},
\end{align}
where $-{\rm T}$ indicates the inverse of the transpose.
Finally, we obtain the ACE operator as
\begin{align}\label{ACE3}
    \hat{V}_\textrm{X}^\textrm{ACE}(\textbf{r},\textbf{r}') = -\sum_{k=1}^{N_e}\zeta_k(\textbf{r})\zeta_k(\textbf{r}').
\end{align}

ACE significantly reduces the cost of application of the exchange operator by reducing the prefactor but keeps the scaling with grid points the same as the scaling for LDA or HF without ACE \cite{lin2016adaptively}. Instead of calculating $\hat{V}_\textrm{X}$ by integration from equation \ref{exchange_op} for each iteration of the SCF cycle, ACE does it only once and improves it iteratively until the SCF density converges. Also, it is important to point out that ACE is not an approximation because after the self-consistency is reached, the action of $\hat{V}_\textrm{X}^\textrm{ACE}$ on an orbital is the same as the action of $\hat{V}_\textrm{X}$. This significantly reduces the overall cost of computation because the cost of application of the exchange operator can often be more than $95\%$ of the total cost in HF calculations.

Once the solution to the ground state is achieved, we calculate a set of unoccupied orbitals via a non-self-consistent calculation. The number of unoccupied orbitals one can obtain depends on the size of the basis set: the number of eigenfunctions obtainable is equal to the size of the Hamiltonian matrix. Hence, the number of unoccupied orbitals obtainable from GTOs will be lower than that obtainable from the (orders of magnitude) larger number of real-space grid points. Usually, one calculates all possible orbitals with GTOs, but in real space only the occupied orbitals (contributing to the Hamiltonian) are calculated in SCF. To generate unoccupied orbitals, a separate non-self-consistent run is performed with a fixed Hamiltonian, calculating only a small fraction of the possible unoccupied orbitals. This is the main difference compared to GTO-based implementations where all possible unoccupied orbitals are usually calculated. Although one can obtain these unoccupied orbitals in the ground-state SCF by calculating an extra set of eigenfunctions with zero occupation which do not contribute to the total energy, it is preferable to calculate the ground state without unoccupied orbitals first, since calculation of these extra eigenfunctions in all iterations before convergence is time-consuming and not useful. Once the number of desired unoccupied orbitals is chosen, the same eigensolver algorithms used for occupied eigenstates in SCF are used to calculate the unoccupied orbitals. Different iterative eigensolvers are available in Octopus such as conjugate gradients \cite{saad_book} and preconditioned Lanczos \cite{saad1996solution}. In this work, we have used conjugate gradients except in some cases where the highest few unoccupied orbitals would not converge. In this situation, continuing with the preconditioned Lanczos algorithm solved the problem. ACE is used here too to calculate the Hamiltonian matrix, with the only difference from the ground-state calculation being the inclusion of unoccupied orbitals in equations \ref{ACE1}, \ref{ACE2} and \ref{ACE3}.

ACE calculates the ground-state exchange potential as a whole iteratively by reducing the number of times the exchange matrix is calculated in the SCF cycle. It is not implemented yet in Octopus to calculate individual exchange integrals with ACE, as would be needed for its use in TDHF. Refs. \cite{sharma2022fast,jones2016efficient} have shown this is possible with plane-waves and Gaussian basis sets using discrete variable representation \cite{colbert1992novel} schemes. Whether this provides any speedup in real space is yet to be studied and is a subject for future work. SCF, without ACE, requires solving $N_e^2$ integrals for each iteration. However, TDA-TDHF requires solving $N_e N_{\rm unocc}$ integrals only once, to create the TDHF matrix which is diagonalized. This means the prefactor for TDHF integrals is generally lower than that for the SCF problem, and so it is more necessary to reduce the cost of exchange in SCF than in TDHF.

\subsection{Mixing in SCF cycle}
The nonlocality of the exchange potential leads to different and more difficult SCF convergence behavior than in semi-local DFT. To facilitate faster convergence, different quantities can be mixed together from consecutive iterations and be used in the next iteration \cite{Castro}. The quantity that is usually mixed is the density, but one can also mix the potential or the wavefunctions. It should also be noted that mixing non-local potentials is not as straightforward as mixing local potentials. In our calculations with potential mixing, we mix only the local part of the potential, i.e. the Hartree potential $V_{\rm H}$.

The simplest mixing scheme is the linear mixing where a certain fraction of the quantity from the last SCF iteration $N-1$ is mixed with that from the current iteration $N$:
\begin{align}\label{linearMixing}
    q_{N+1}(\textbf{r}) = \alpha q_{N-1}(\textbf{r}) + (1-\alpha) q_{N}(\textbf{r})
\end{align}
where $q$ refers to the quantity that is being mixed and $\alpha$ is called the ``mixing parameter.''
More sophisticated schemes like Broyden mixing \cite{broyden1965class, johnson1988modified} and Pulay mixing \cite{pulay1980convergence} are widely used too. 

\subsection{Pseudopotentials}\label{PPs}
In real-space or plane-wave calculations, pseudopotentials (PPs) are the standard approach to account only implicitly for the core electrons of a system, which can be done because the core electrons usually do not take part in bonding or electronic and optical properties. In real space, PPs also solve a crucial problem. The $-Z/r$ Coulomb potential is divergent at $r=0$ and makes the wavefunctions vary very rapidly near the nucleus, requiring an infeasibly large number of grid points to describe them. PPs circumvent this problem \cite{kronik2006parsec}. By contrast, in Gaussian basis sets, usually all the electrons are treated explicitly, except for heavy elements in which case effective core potentials (a form of PPs) \cite{cundari1993effective} are used. PPs are generated from all-electron calculations for a given DFT functional. Typically only local or semi-local functionals are used. Approaches to generate PPs for HF have been developed \cite{trail2005norm,burkatzki2007energy} but such PPs are rarely used in practice \cite{tan2018effect} and are not supported by Octopus. In this paper, we use optimized norm-conserving Vanderbilt \cite{Hamann2013} scalar relativistic pseudopotentials from pseudo-dojo \cite{PseudoDojo}. We use LDA \cite{PhysRevB.45.13244, dirac1930note} pseudopotentials for all HF ground-state calculations, as well as the LDA calculations reported later. For DFT calculations with BLYP \cite{Becke88,LYP} for our TDDFT comparison, we have used PBE PPs as BLYP and PBE are both GGA functionals and pseudo-dojo currently doesn't have BLYP PPs. For these TDDFT calculations, we note that Octopus currently only supports LDA kernels and so the LDA kernel \cite{PhysRevB.45.13244, dirac1930note} is used for our TDDFT calculations based on a BLYP ground state, as is often done in Octopus calculations \cite{Vila}. 
Using PPs generated with one functional for a DFT calculation with a different functional is a common practice in DFT, and ubiquitous for hybrid functionals. A recent study shows that the inconsistent use of PPs does not make very significant differences \cite{borlido2020validation} in DFT.

\subsection{Real-space calculations}\label{globalPara}
In the Octopus code \cite{andrade2012time,andrade2015real,Octopus_2020}, all quantities are represented on a grid in real space. The domain is divided into a grid, with zero boundary conditions applied on the surface. The default (``minimal'') simulation domain for molecules is a union of spheres centered around each atom.
However, for the results presented in this paper, we have used a single big sphere that encapsulates the entire molecule. For all the molecules used in this work, the atoms would fit in a sphere of radius $R = 3$ \r{A}, so for sphere radii much larger than that, the union-of-spheres domain is not very different from a single sphere. The radius of the encapsulating sphere used for benchmarking results is $10$ \r{A} and a spacing $h = 0.1$ \r{A} is used. These parameters, which define the real-space grid, can be converged systematically. In our case, they were selected by convergence tests of not only occupied but also unoccupied orbitals, as shown in section \ref{conv_orbitals}, since convergence of unoccupied orbitals is very important for excited-state calculations. 
We use 51 unoccupied orbitals for each molecule. The conjugate-gradients eigensolver was used, with a tolerance of $10^{-6}$ eV. 
By default, Octopus uses as SCF convergence criterion the relative density error which measures the relative change of electron density between iterations as given by the following equation:
\begin{align}
    \epsilon = \frac{1}{N_e}\int\left| \rho^{\rm in} \left( \textbf{r} \right) -\rho^{\rm out} \left( \textbf{r} \right) \right| d^3\textbf{r}
    \label{reldens}
\end{align}
where $\rho^{\rm in}$ and $\rho^{\rm out}$ are the densities at two consecutive iterations.
Requiring eigensolver convergence for each state as an additional convergence criterion in SCF led to density convergence several orders of magnitude better than the default threshold $10^{-6}$ for the relative density error, which is related to the fact that the HF Hamiltonian depends directly on the wavefunctions, not just the density.

The workflow of TDHF in Octopus is very similar to that of a standard linear-response TDDFT calculation \cite{tutorial}. With our implementation, we can now use flag {\tt CalculationMode = casida} along with {\tt TheoryLevel = hartree\_fock}. For the TDA calculations shown in this study, an additional flag, {\tt CasidaTheoryLevel = tamm\_dancoff}, has been used. The use of ACE in the ground state and unoccupied orbitals calculations can be turned on via flag {\tt AdaptivelyCompressedExchange = yes}.

\section{Results}\label{results}
\subsection{Speed-up of HF with ACE}
\begin{figure}[H]
    \centering
    \includegraphics[width=\textwidth]{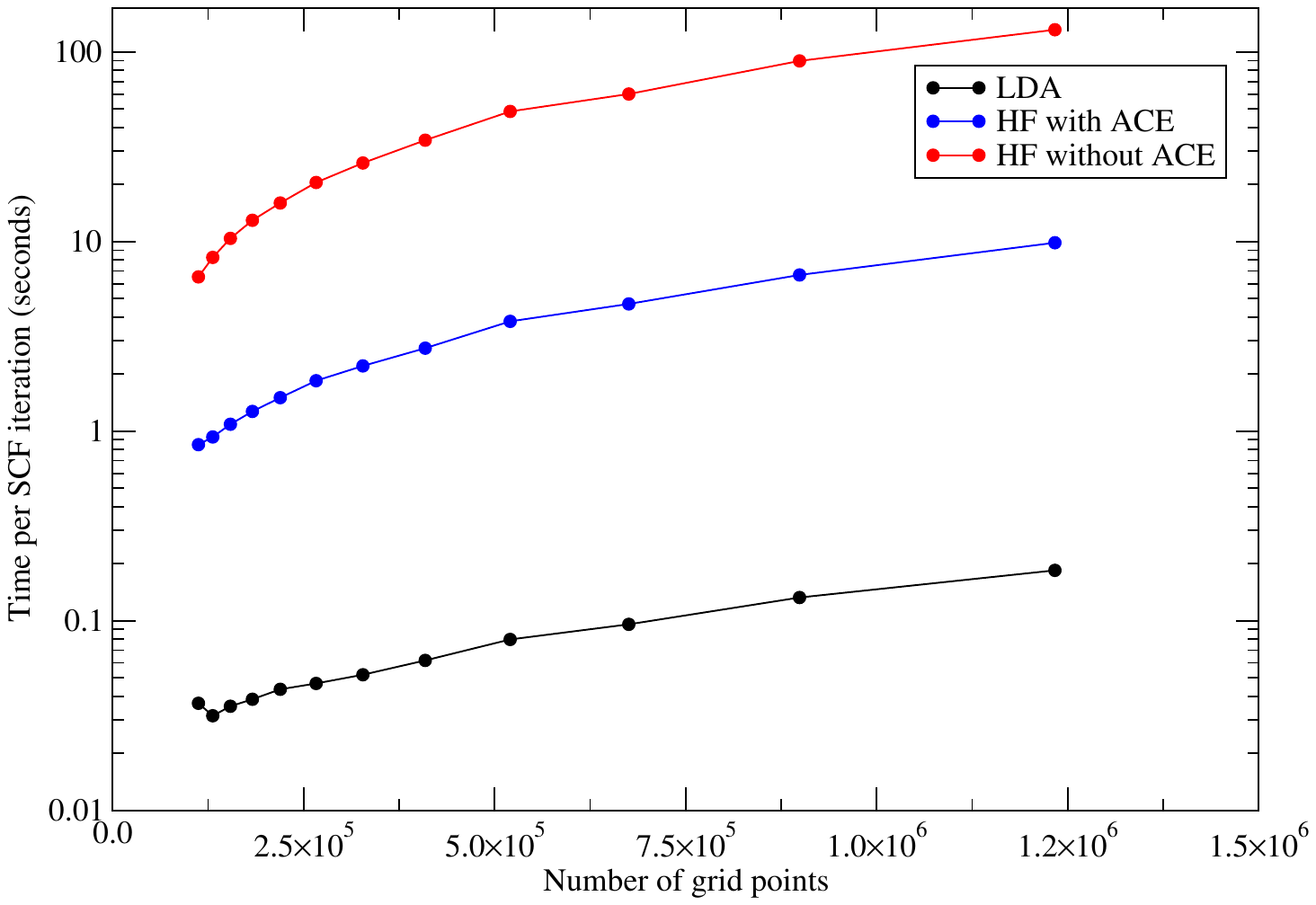}
    \caption{Scaling of time per SCF iteration for different approaches, showing similar scaling for each but an improved pre-factor for HF when using ACE.  Times are calculated as an average for the first 20 SCF iterations for the nitrosomethane molecule. Data was obtained for a fixed number of orbitals by varying the grid spacing ($h$) by $0.01$ \r{A} from $0.20$ \r{A} to $0.09$ \r{A} on a simulation domain constructed from a union of spheres of radii $5$ \r{A} around each atom.}
    \label{ACE_scaling}
\end{figure}

To test the efficiency of the ACE formalism, we can look at the average time required to complete one iteration in the SCF cycle. Table \ref{TimePerIteration} shows that the time per iteration for LDA calculations is $1-3\%$ of the time per iteration for HF calculations without ACE, i.e. the direct solution of the Poisson equations for exchange. With ACE, however, we have achieved a speed-up of at least 7 times, with maximum speed-up of 14 times for formamide. We have also confirmed numerically identical results between calculations with and without ACE. We also studied the scaling of time per SCF iteration as a function of number of grid-points for LDA, HF without ACE, and HF with ACE, as shown in Figure \ref{ACE_scaling}. We see similar scaling in each case: the scaling should be $\mathcal{O} \left( N_\mu \right)$ for LDA and $\mathcal{O} \left( N_\mu \log N_\mu \right)$ for HF, which look similar on this scale. However, the speed-up due to ACE is dramatic. This is a big improvement in making the ground-state HF calculations practical: without ACE they were two orders of magnitude more expensive than LDA, but with ACE are only one order of magnitude more expensive.

\begin{table}[H]
\small
\begin{tabular}{|l||c||ccc|}
\hline
\multirow{2}{*}{Molecule} & \multirow{2}{*}{\begin{tabular}[c]{@{}c@{}}Time per SCF iteration (seconds)\\ (DFT-LDA)\end{tabular}} & \multicolumn{3}{c|}{\begin{tabular}[c]{@{}c@{}}Time per SCF iteration (seconds)\\ (HF)\end{tabular}} \\ \cline{3-5} 
                          &                                                                                             & \multicolumn{1}{c|}{Without ACE}    & \multicolumn{1}{c|}{With ACE}    & Speed-up    \\ \hline \hline
acetaldehyde              & 0.19                                                                                        & \multicolumn{1}{c|}{20.27}       & \multicolumn{1}{c|}{1.71}           & 11.85             \\ \hline
acetylene                 & 0.05                                                                                        & \multicolumn{1}{c|}{3.41}        & \multicolumn{1}{c|}{0.42}           & 8.14              \\ \hline
ammonia                   & 0.05                                                                                        & \multicolumn{1}{c|}{2.12}        & \multicolumn{1}{c|}{0.28}           & 7.59              \\ \hline
carbon monoxide           & 0.05                                                                                        & \multicolumn{1}{c|}{2.92}        & \multicolumn{1}{c|}{0.34}           & 8.58              \\ \hline
cyclopropene              & 0.07                                                                                        & \multicolumn{1}{c|}{12.33}       & \multicolumn{1}{c|}{1.35}           & 9.14              \\ \hline
diazomethane              & 0.08                                                                                        & \multicolumn{1}{c|}{11.12}       & \multicolumn{1}{c|}{1.12}           & 9.93              \\ \hline
dinitrogen                & 0.05                                                                                        & \multicolumn{1}{c|}{2.57}        & \multicolumn{1}{c|}{0.34}           & 7.68              \\ \hline
ethylene                  & 0.05                                                                                        & \multicolumn{1}{c|}{5.12}        & \multicolumn{1}{c|}{0.66}           & 7.76              \\ \hline
formaldehyde              & 0.12                                                                                        & \multicolumn{1}{c|}{10.62}       & \multicolumn{1}{c|}{1.09}           & 9.71              \\ \hline
formamide                 & 0.09                                                                                        & \multicolumn{1}{c|}{22.01}       & \multicolumn{1}{c|}{1.57}           & 14.02             \\ \hline
hydrogen chloride         & 0.04                                                                                        & \multicolumn{1}{c|}{1.88}        & \multicolumn{1}{c|}{0.26}           & 7.33              \\ \hline
hydrogen sulphide         & 0.05                                                                                        & \multicolumn{1}{c|}{2.08}        & \multicolumn{1}{c|}{0.27}           & 7.68              \\ \hline
ketene                    & 0.07                                                                                        & \multicolumn{1}{c|}{11.71}       & \multicolumn{1}{c|}{1.13}           & 10.38             \\ \hline
methanimine               & 0.06                                                                                        & \multicolumn{1}{c|}{6.54}        & \multicolumn{1}{c|}{0.67}           & 9.82              \\ \hline
nitrosomethane            & 0.07                                                                                        & \multicolumn{1}{c|}{22.34}       & \multicolumn{1}{c|}{1.64}           & 13.61             \\ \hline
streptocyanine-c1         & 0.10                                                                                        & \multicolumn{1}{c|}{20.88}       & \multicolumn{1}{c|}{1.97}           & 10.63             \\ \hline
thioformaldehyde          & 0.06                                                                                        & \multicolumn{1}{c|}{6.25}        & \multicolumn{1}{c|}{0.66}           & 9.46              \\ \hline
water                     & 0.04                                                                                        & \multicolumn{1}{c|}{1.84}        & \multicolumn{1}{c|}{0.25}           & 7.46              \\ \hline
\end{tabular}
\caption{Time per SCF iteration in different molecules, for runs in parallel using a node with two Intel 28 core Xeon Gold 6330 processors. Times are calculated by taking an average over the first 10 iterations for HF and an average over all iterations for LDA.}
\label{TimePerIteration}
\end{table}

\subsection{Starting point for SCF convergence of HF}
The ground-state SCF calculation can be difficult to converge for HF. The default initial guess in Octopus is a linear combination of atomic orbitals (LCAO) for the wavefunctions and a superposition of atomic densities for the density. One way to accelerate this convergence is to first calculate the ground state using LDA (faster and more easily converged) and then use that a starting point instead of using LCAO calculated in HF as the initial guess for the HF calculation, assuming that the converged density will be similar. This approach is the standard recommendation in Octopus for hybrid functionals. To assess the convergence behaviour, we examine the relative density error as defined in equation \ref{reldens}.

\begin{figure}[H]
    \centering
    \includegraphics[width=\textwidth]{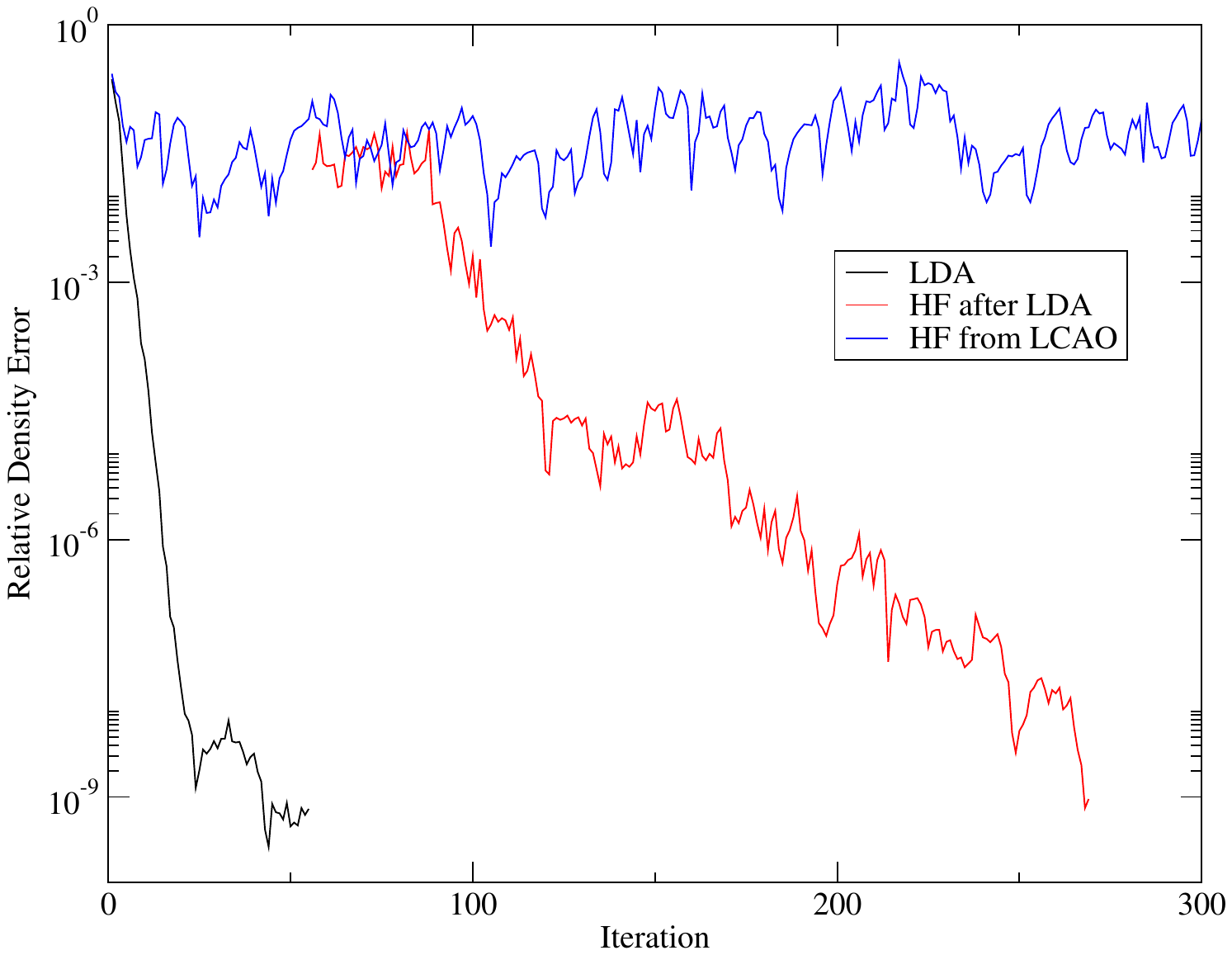}
    \caption{Convergence of HF ground state for thioformaldehyde ($h=0.22$ \r{A}, $R=5$ \r{A}) from different starting points, measured by the relative density error defined in equation \ref{reldens}. First an LDA calculation (black) was performed starting from a Linear Combination of Atomic Orbitals (LCAO) guess and the converged results were used as initial guess for HF calculation (red). The blue curve shows convergence when the HF calculation was performed directly from LCAO.}
    \label{convergenceLDAHF}
\end{figure}

Figure \ref{convergenceLDAHF} shows the convergence characteristics of the SCF procedure for different starting guesses using Broyden density mixing. 
It is evident from Figure \ref{convergenceLDAHF} that starting from the LDA solution not only makes the convergence faster but is also responsible for any convergence at all, as the HF from LCAO didn't converge within 300 iterations and was not making progress. 

\subsection{Mixing for SCF convergence of HF}

\begin{figure}
  \centering
  \begin{tikzpicture}
    \node at (-1.55,0) {\includegraphics[width=0.5\textwidth]{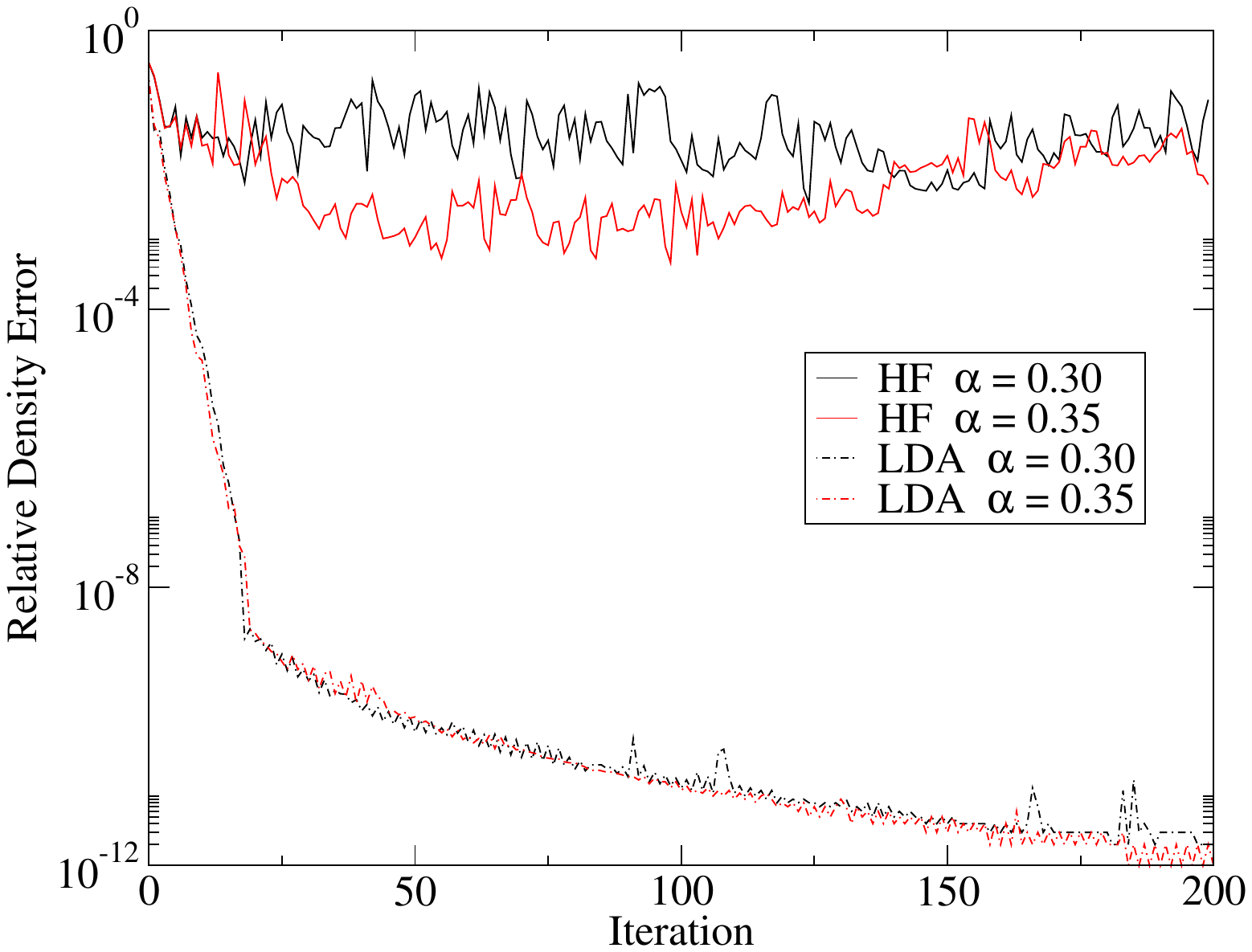}};
    \node at (-.7,0.2\textwidth) {a) Broyden potential mixing};

    \node at (0.4\textwidth,0) {\includegraphics[width=0.5\textwidth]{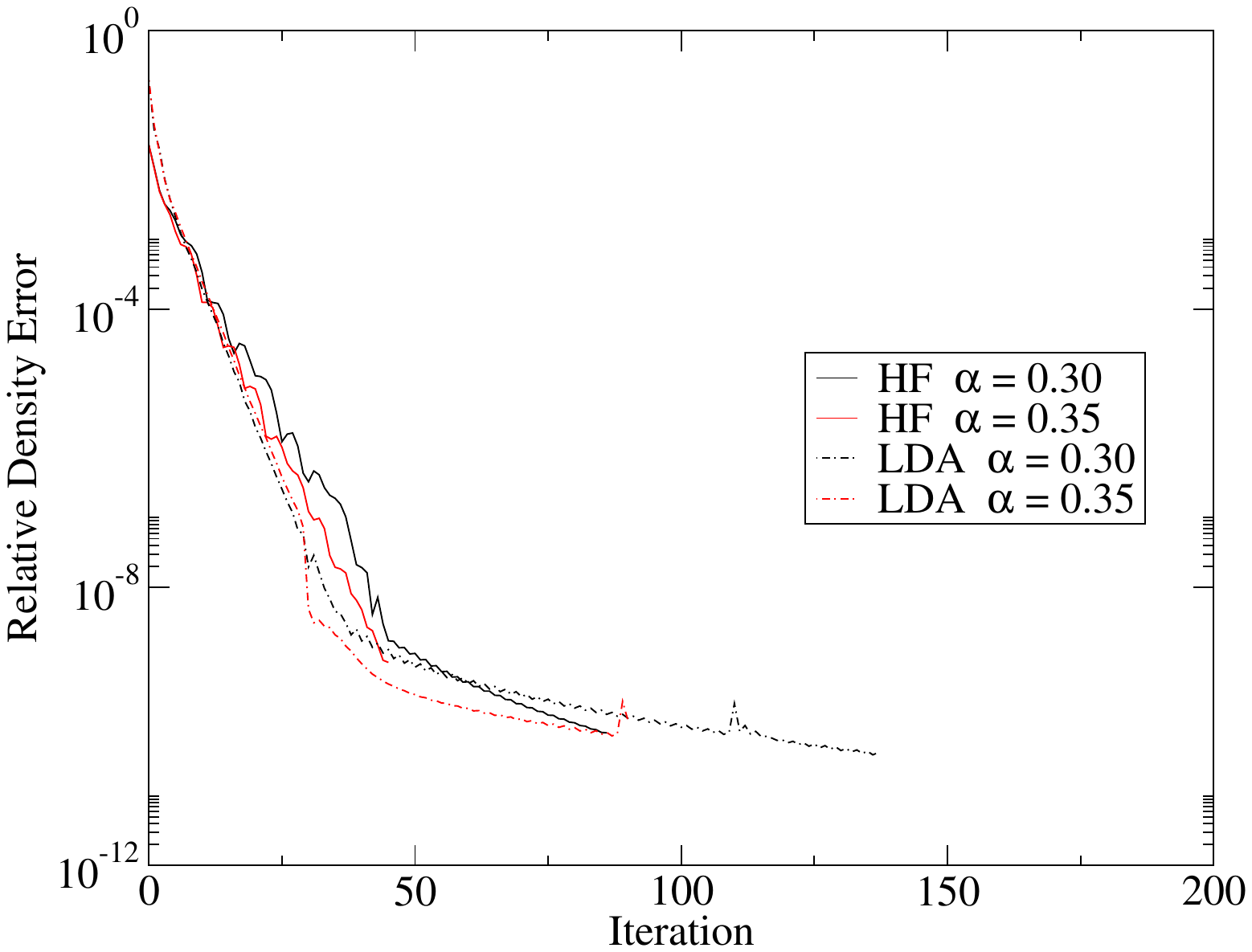}};
    \node at (0.4\textwidth,0.2\textwidth) {b) Linear potential mixing};

    \node at (-1.55,-0.4\textwidth) {\includegraphics[width=0.5\textwidth]{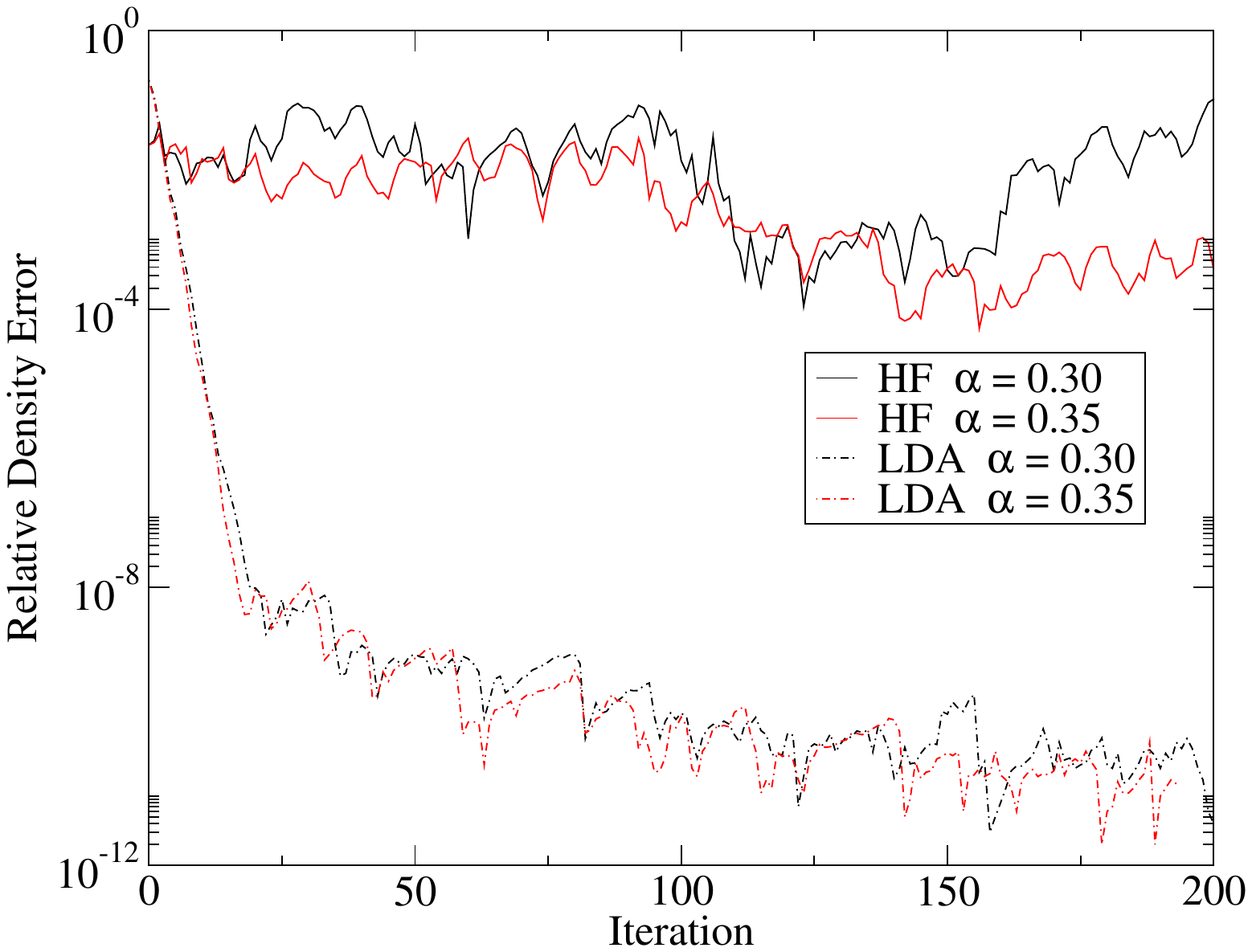}};
    \node at (-0.7,-0.2\textwidth) {c) Broyden density mixing};

    \node at (0.4\textwidth,-0.4\textwidth) {\includegraphics[width=0.5\textwidth]{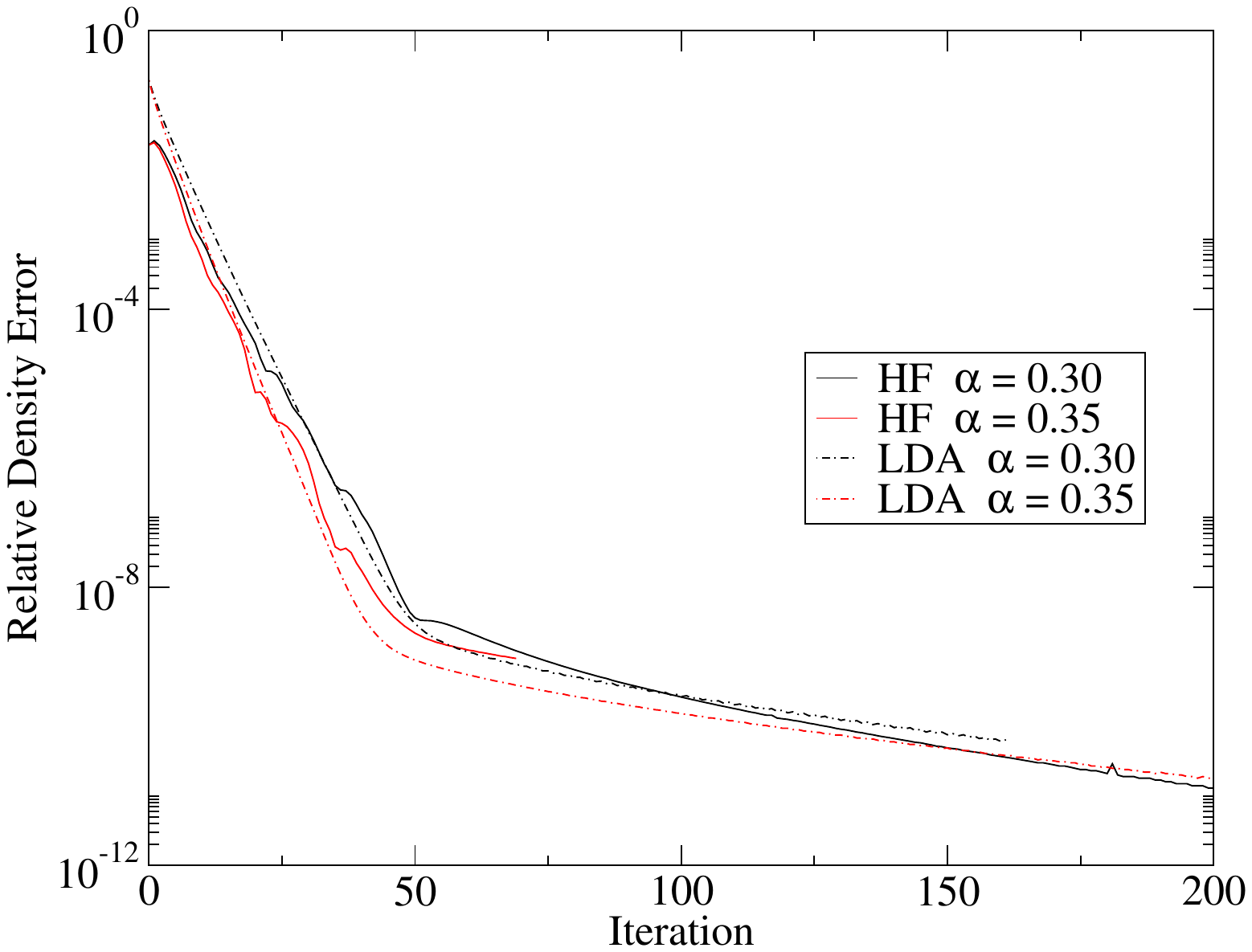}};
    \node at (0.4\textwidth,-0.2\textwidth) {d) Linear density mixing};
  \end{tikzpicture}
  \caption{Effect of the mixing scheme, quantity being mixed, and the mixing parameter $\alpha$ on convergence. The relative density error is defined in equation \ref{reldens}.}
  \label{Mixing}
\end{figure}

\begin{figure}[H]
    \centering
    \includegraphics[width=.5\textwidth]{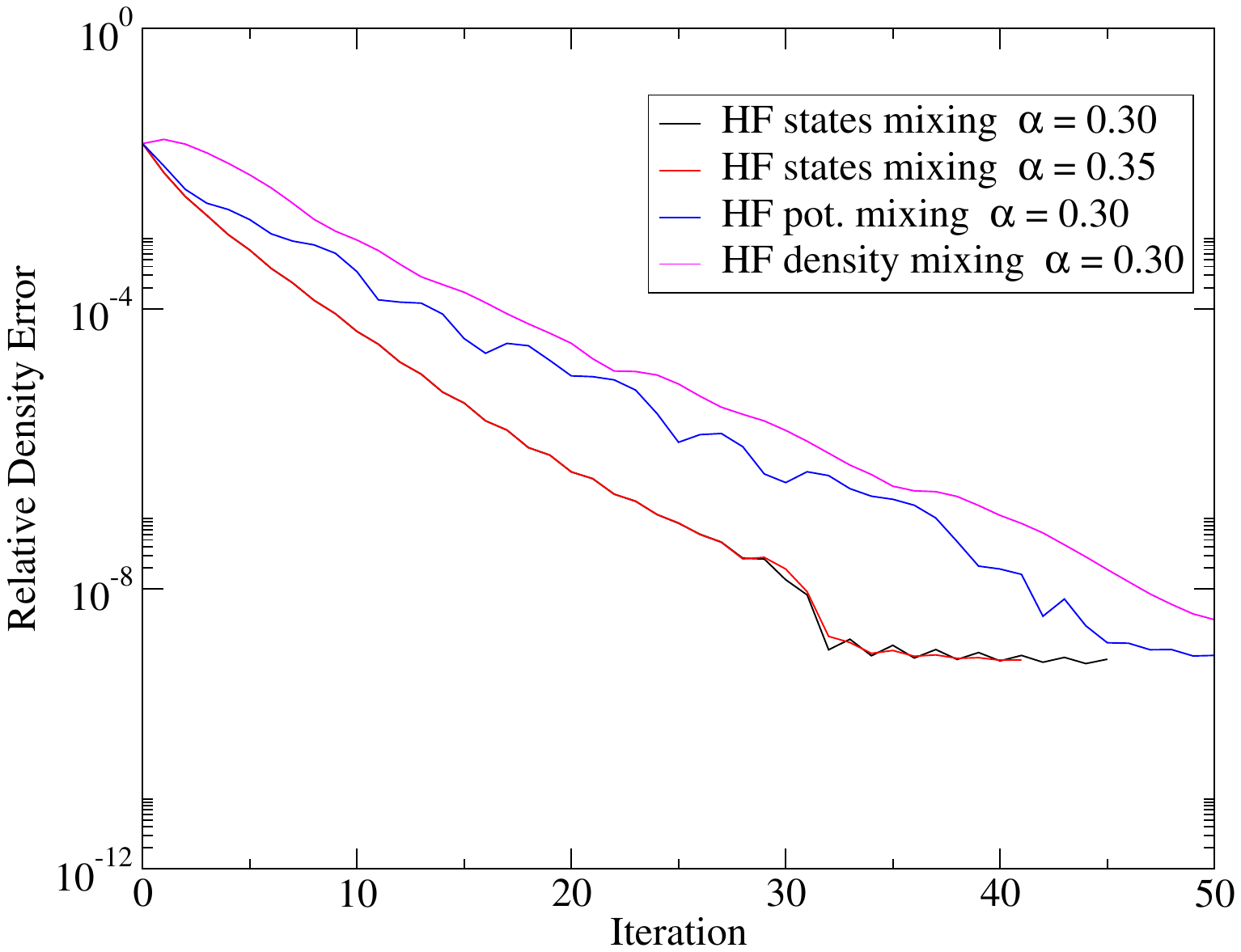}
    \caption{Convergence, as measured by the relative density error defined in equation \ref{reldens}, while mixing states (red and black with different mixing parameters) using the linear scheme for the nitrosomethane molecule ($h=0.16$\ \r{A}, $R=5$\ \r{A}). Blue and magenta curves are linear mixing of potential and density, respectively, shown as a comparison to states mixing.}
    \label{MixStates}
\end{figure}

Figure \ref{Mixing} shows the convergence characteristics using different mixing quantities and mixing schemes for nitrosomethane at mixing parameters $\alpha$ = $0.30$ and $0.35$. All HF calculations start using LDA solutions as initial guess. Although for LDA the Broyden scheme seems faster and smoother than linear mixing, HF doesn't converge at all with this scheme and linear mixing should be used to achieve convergence. 
For LDA, Broyden mixing of potential and density have similar behavior except that potential mixing provides smoother convergence at very low relative density error. For both LDA and HF, linear mixing of potential and density have similar behavior. The differences in all cases between our two values of $\alpha$ is small whenever convergence is achieved.
We also show the convergence character using linear mixing of states for the same system in Figure \ref{MixStates}, showing significantly improved behavior. Mixing states seems the most robust strategy for faster convergence of SCF for a variety of molecules. This makes sense because the HF potential is not just a function of density but also of the orbitals, and converging orbitals should guarantee the convergence of both potential and density.

\subsection{Convergence of Orbitals with the Real-Space Grid Parameters}\label{conv_orbitals}
\begin{figure}
  \centering
\begin{tikzpicture}
    \node at (-2,0) {\includegraphics[width=.55\textwidth]{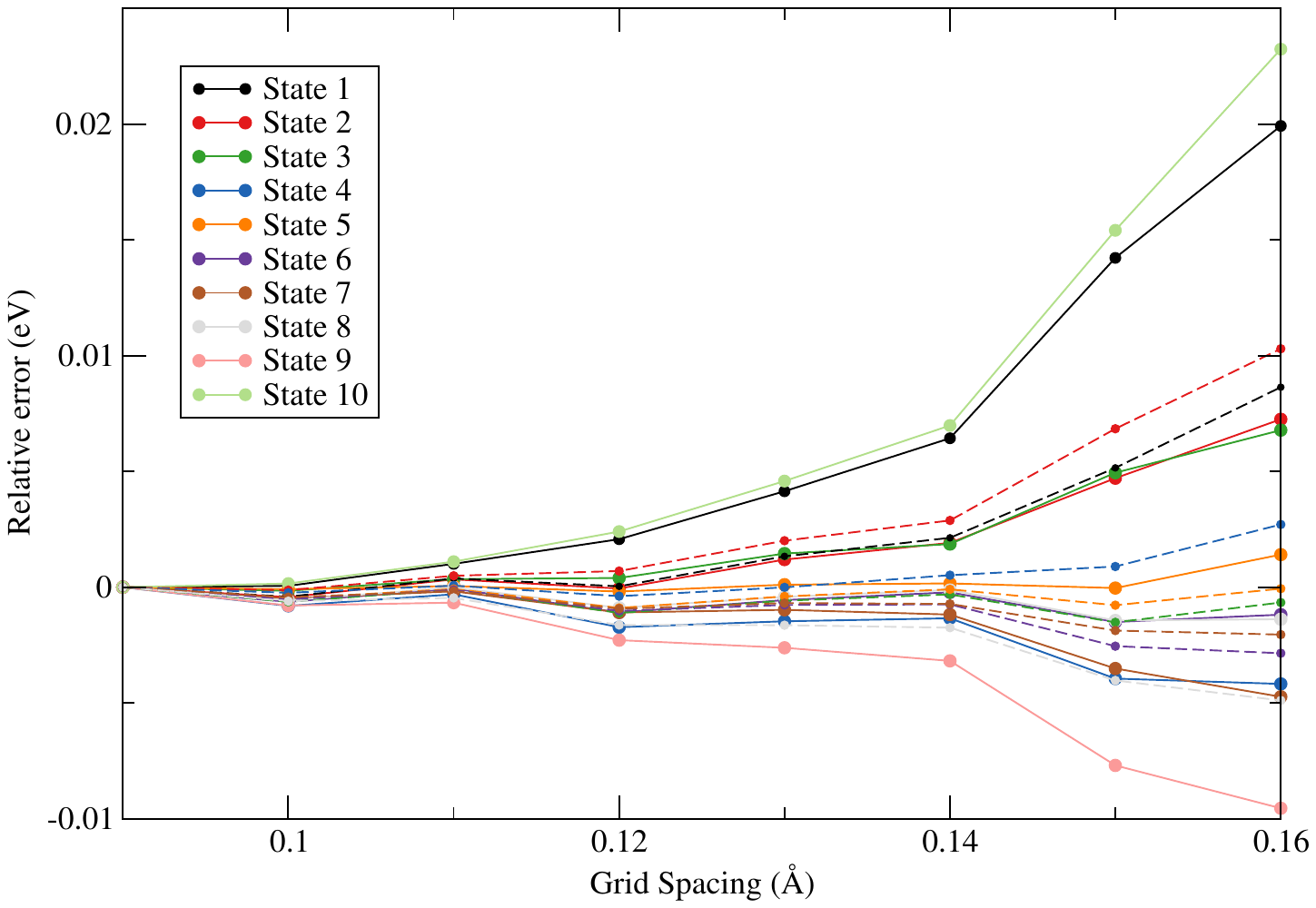}};
    \node at (-.12\textwidth,0.17\textwidth) {a) Occupied};

    \node at (0.38\textwidth,0) {\includegraphics[width=0.55\textwidth]{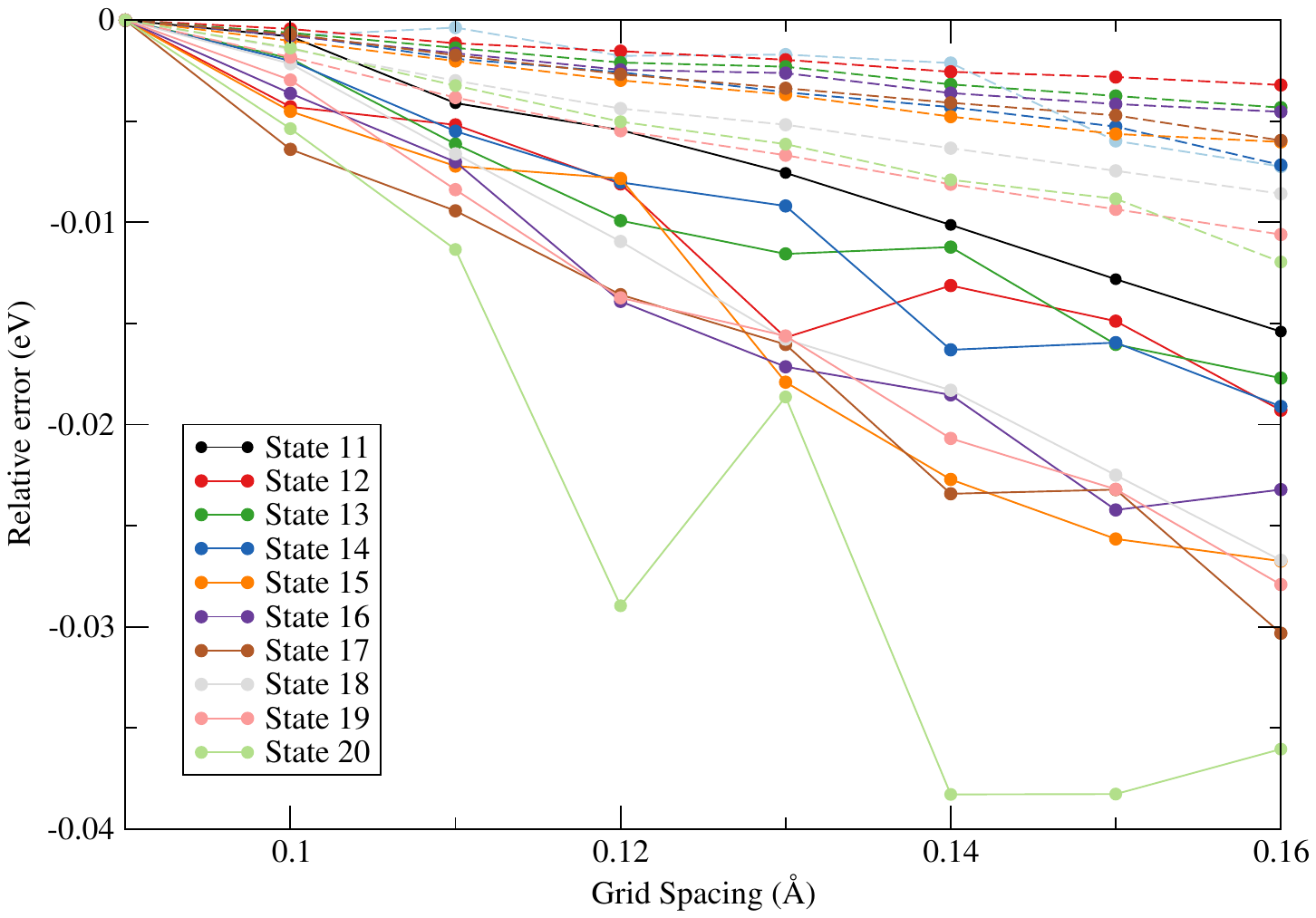}};
    \node at (0.4\textwidth,0.17\textwidth) {b) Unoccupied};

\end{tikzpicture}
\caption{Relative error of the occupied and unoccupied orbitals' eigenvalues compared to the finest grid ($h=0.09$ \r{A}) at various grid spacings for thioformaldehyde  on a domain of union of spheres of radii $5$ \r{A}. Dashed and solid lines represent LDA and HF respectively.}
\label{convEval}
\end{figure}

To obtain TDHF spectra converged with respect to the grid spacing $h$ and sphere radius $R$, we must converge not only the ground-state density, but also the occupied and unoccupied orbitals' eigenvalues and wavefunctions.
We show the convergence of occupied and unoccupied eigenvalues in figure \ref{convEval} from HF and LDA with respect to the change in the grid spacing of the calculation domain.

\begin{figure}[H]
    \centering
    \includegraphics[width=.7\textwidth]{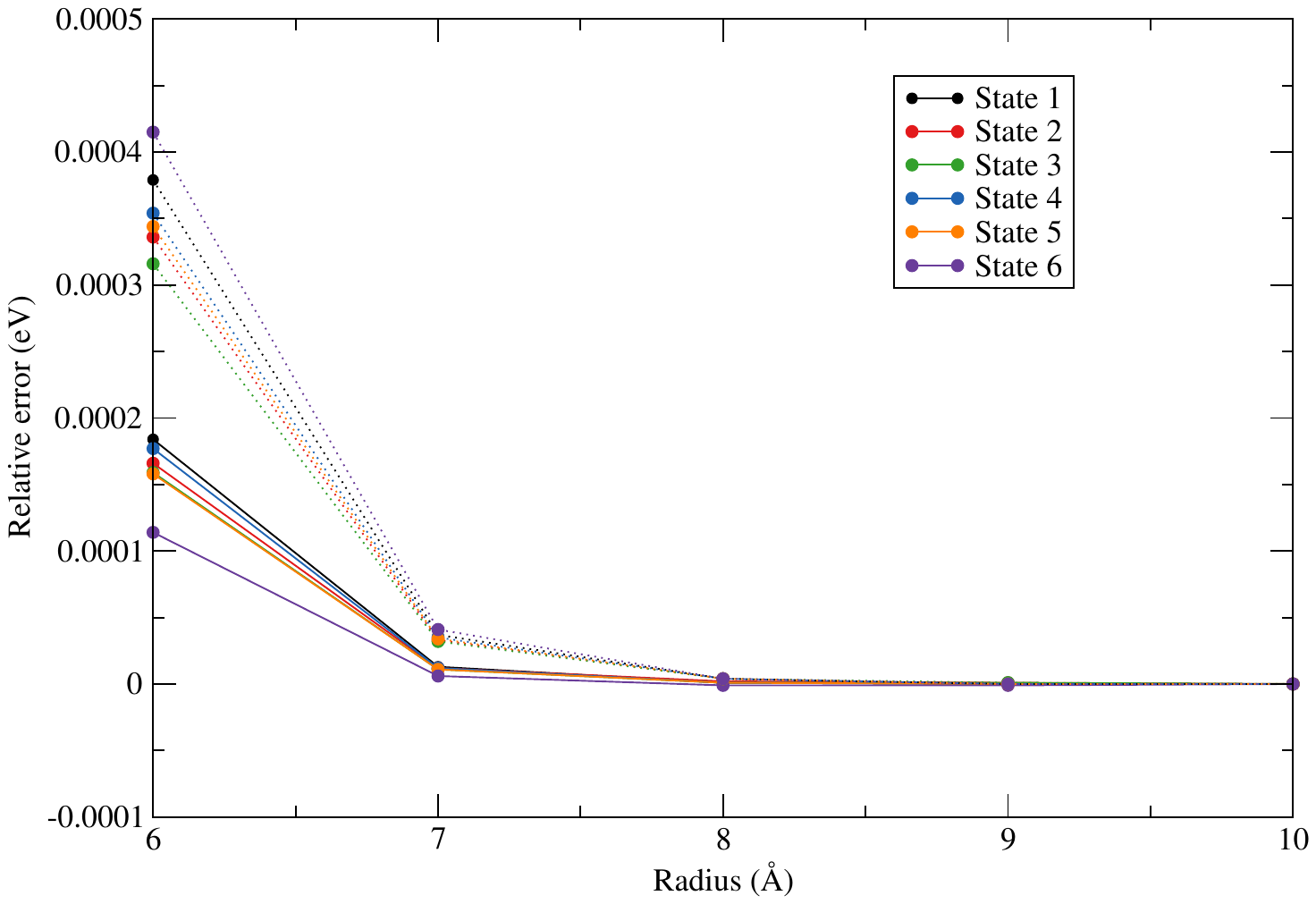}
    \caption{Relative error of the occupied orbitals' eigenvalues compared to the biggest simulation sphere ($R=10$\ \r{A}) at various radii for ethylene molecule with grid spacing $h = 0.08$ \r{A}. Dashed and solid lines represent  LDA and HF respectively.}
    \label{convOccEvalwRadius}
\end{figure}

The occupied orbitals %
converge within $0.01$ eV at $0.14$ \r{A} for both LDA and HF, and have similar convergence behavior.
The unoccupied orbitals similarly converge within $0.01$ eV at $0.14$ \r{A} for LDA, but for HF a finer grid of spacing $0.1$ \r{A} is required to achieve the same level of convergence. This is not surprising since there is an additional integral to be converged in real space, in HF vs LDA. The occupied orbitals converge monotonically as the spacing is decreased in both LDA and HF. For unoccupied orbitals, convergence is monotonic for LDA too, but not for HF. 
In Figure \ref{convOccEvalwRadius} we examine also the convergence of occupied eigenvalues with respect to the sphere radius $R$, which shows that eigenvalues are very well converged by $R = 10$\ \AA. This is as expected since that large radius was chosen to ensure adequate description of the unoccupied orbitals. It is not straightforward to make a meaningful assessment of unoccupied eigenvalues' convergence with radius, because the state of a given index will continually decrease toward zero as the radius is increased, and converge only to zero. These vacuum states behave similarly to the particle-in-a-box model. Therefore we do not provide a plot of unoccupied states' convergence with radius.
Overall we conclude that convergence of HF unoccupied orbitals is harder in real space but still achievable.

\subsection{Benchmarking TDHF}\label{Benchmarking}

For benchmarking purposes, we've chosen to compare our real-space excitation energy calculations in Octopus to those from \cite{gould2022single} which uses the GTO code Psi4 \cite{parrish2017psi4}. Out of the $17$ molecules from the GTO dataset studied there (ultimately from \cite{loos2018mountaineering}, including geometries), we excluded formaldehyde, methamine and streptocyanine-c1 from our comparison due to problems with convergence of unoccupied levels in real space. We have compared the excitation energies obtained from these two different codes for IPA and TDA from TDHF as well as from TDDFT (using the BLYP functional).

\begin{figure}
  \centering
\begin{tikzpicture}
    \node at (0,0) {\includegraphics[width=0.3\textwidth]{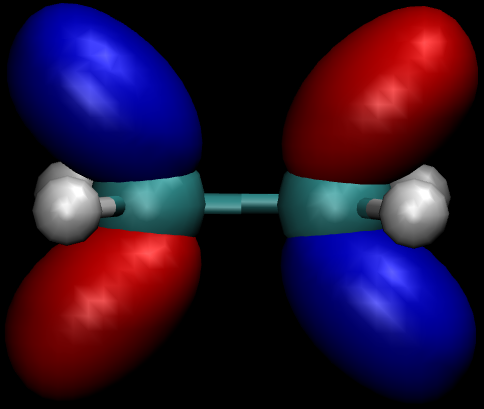}};
    \node at (0,0.15\textwidth) {a) GTO Valence LUMO+4 at $2.50$ eV};

    \node at (0.45\textwidth,0) {\includegraphics[width=0.25\textwidth]{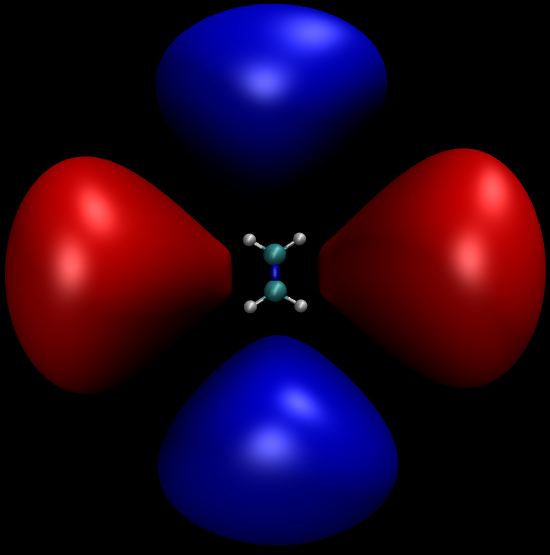}};
    \node at (0.45\textwidth,0.15\textwidth) {b) Real-space Rydberg LUMO+4 at $1.25$ eV};

    \node at (0,-0.3\textwidth) {\includegraphics[width=0.3\textwidth]{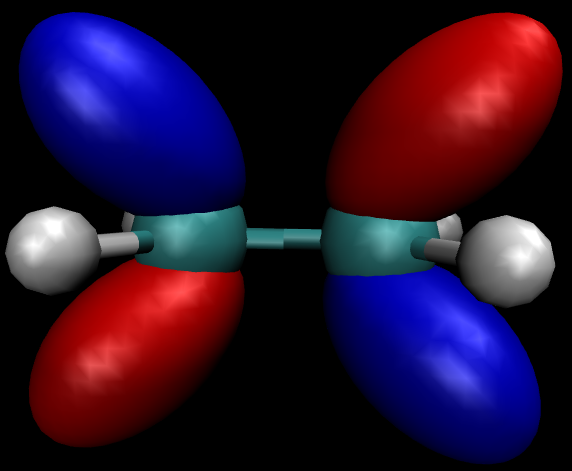}};
    \node at (0,-0.16\textwidth) {c) Real-space Valence LUMO+29 at 2.82 eV};

\end{tikzpicture}
\caption{Unoccupied orbitals of ethylene molecule. GTO-based orbital is obtained from Psi4 using aug-cc-pvtz basis set. Real-space orbitals are obtained from Octopus in a simulation sphere with $R = 10$ \r{A} and $h = 0.1$ \r{A}. }
\label{Ethylene_orbitals}
\end{figure}

The unoccupied  orbitals and their symmetry is sensitive to the choice of basis set. The number of virtual orbitals obtainable depends on the number of basis functions in GTOs while in real-space it depends on the number of grid-points. For example, using the aug-cc-pvtz basis set for acetaldehyde gives 496 basis functions \cite{gould2022single}, meaning a total of 496 eigenstates can be calculated. But in real space we are not limited to that number and finer grids give us a larger number of eigenstates. The extra states obtained in real-space calculations are mostly low-lying Rydberg states. This makes the process of targeting the same excitation for comparison in GTO and real-space-based approaches difficult. The indices of the states targeted in \cite{gould2022single} are different from those of the corresponding states in a real-space calculation. The size of the simulation domain also affects the nature of the calculated orbitals. Increasing the domain size results in introduction of new low-lying vacuum states which may not contribute to the absorption spectrum but nonetheless are important in understanding other kinds of response \cite{Vila}.  Figure \ref{Ethylene_orbitals}a) shows the state targeted for the triplet excitation as obtained using GTOs, which is the fourth state after the lowest unoccupied molecular orbital (LUMO). However, in real space, the same state index represents a Rydberg orbital as shown in figure \ref{Ethylene_orbitals}b). The state resembling the GTO LUMO+4 state is found in real space as the $30^{\rm th}$ unoccupied state. With this issue in mind, for the comparison of excitation energies against the data from \cite{gould2022single}, we have ensured that we are targeting the same excitation from both approaches by making sure the symmetry of the unoccupied orbitals matches from both approaches along with their IPA excitation energies. It is harder to determine  the symmetry of the orbitals in real space. The symmetry of GTOs can be used to classify orbital symmetries, typically in the solution process. In real space, the Cartesian grid inherently breaks the system's symmetry in most cases. %
In this work, we have manually examined the molecular orbitals to determine their irreducible representations.

Table \ref{RSvsGTOs_Errors} summarizes the mean error (ME) and the mean absolute error (MAE) of RS excitation energies as compared to the GTO approach from both TDDFT and TDHF for singlets and triplets. The MAE for TDHF is 2 times larger than TDDFT for singlets and 5 times for triplets, showing that exact-exchange integrals introduce a  greater discrepancy. Triplets have a larger MAE than singlets, which we attribute in part to favorable cancellation of errors for the singlet between the two integrals of equation \ref{RHF_matrix}, whereas only one term is present for the triplet. MAEs are about an order of magnitude bigger for HF eigenvalues (IPA) than for BLYP, and HF eigenvalues' MAE is 50-100\% of the TDHF MAE. This comparison indicates that the eigenvalues, coming from the long-standing and well-tested HF implementation, rather than the TDHF integrals implemented in this work, are responsible for a large fraction of the discrepancies for TDHF. These discrepancies therefore are considered to stem from the inherent numerical differences between the RS and GTO calculations.
In tables \ref{Bmark_trips} and \ref{Bmark_sings}, we show all the triplet and singlet excitation energies, respectively, as obtained from TDDFT (TDBLYP) and TDHF, in the IPA and TDA forms, from both the GTO-based Psi4 code and the real-space (RS) Octopus code. 
The TBE column represents the ``Theoretical Best Estimate'' from GTO coupled-cluster calculations in \cite{loos2018mountaineering}.

\begin{figure}
  \centering
\begin{tikzpicture}
    \node at (-0.2,0) {\includegraphics[width=.7\textwidth]{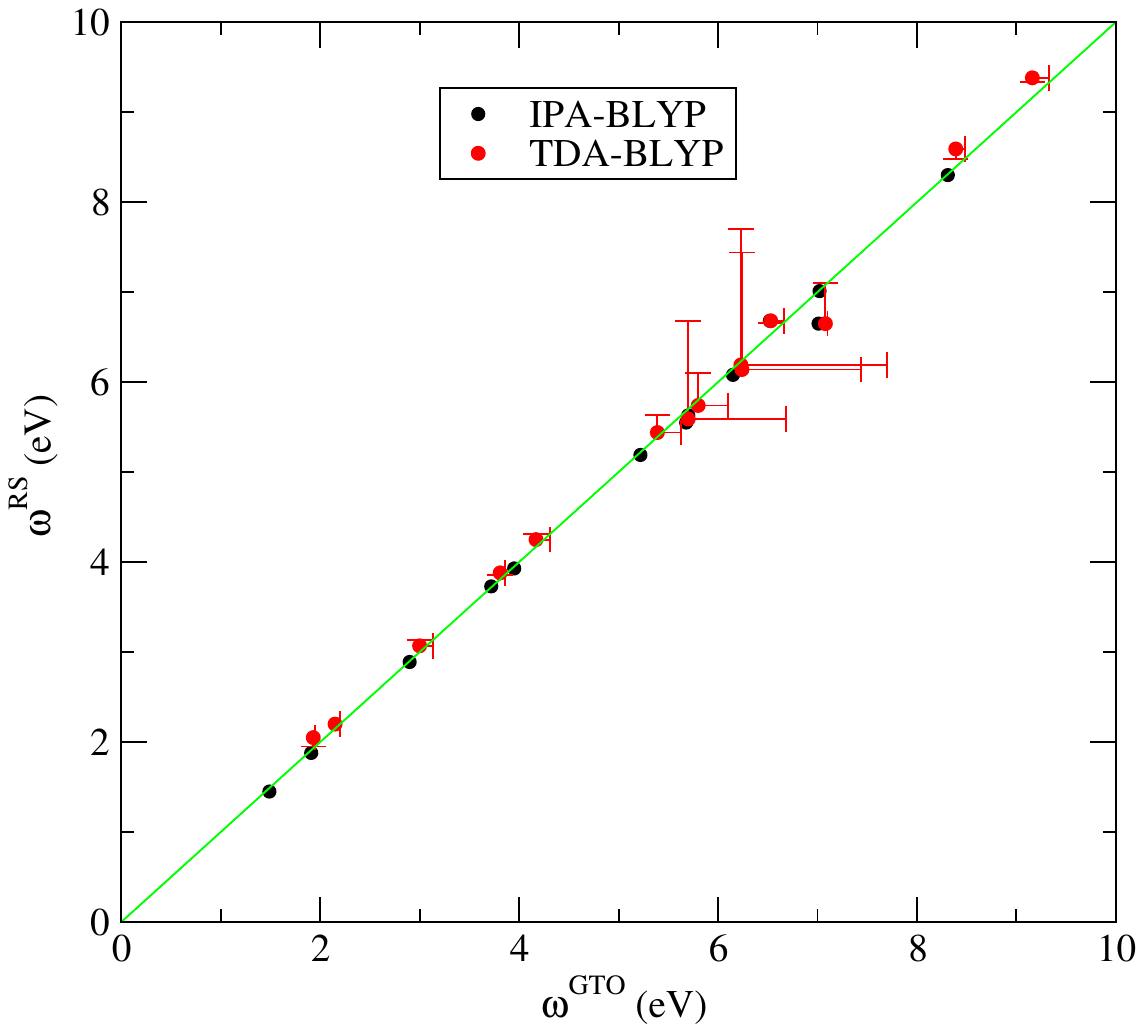}};
    \node at (-0.3,0.21\textwidth) {a) Singlets};

    \node at (0.5\textwidth,0) {\includegraphics[width=0.7\textwidth]{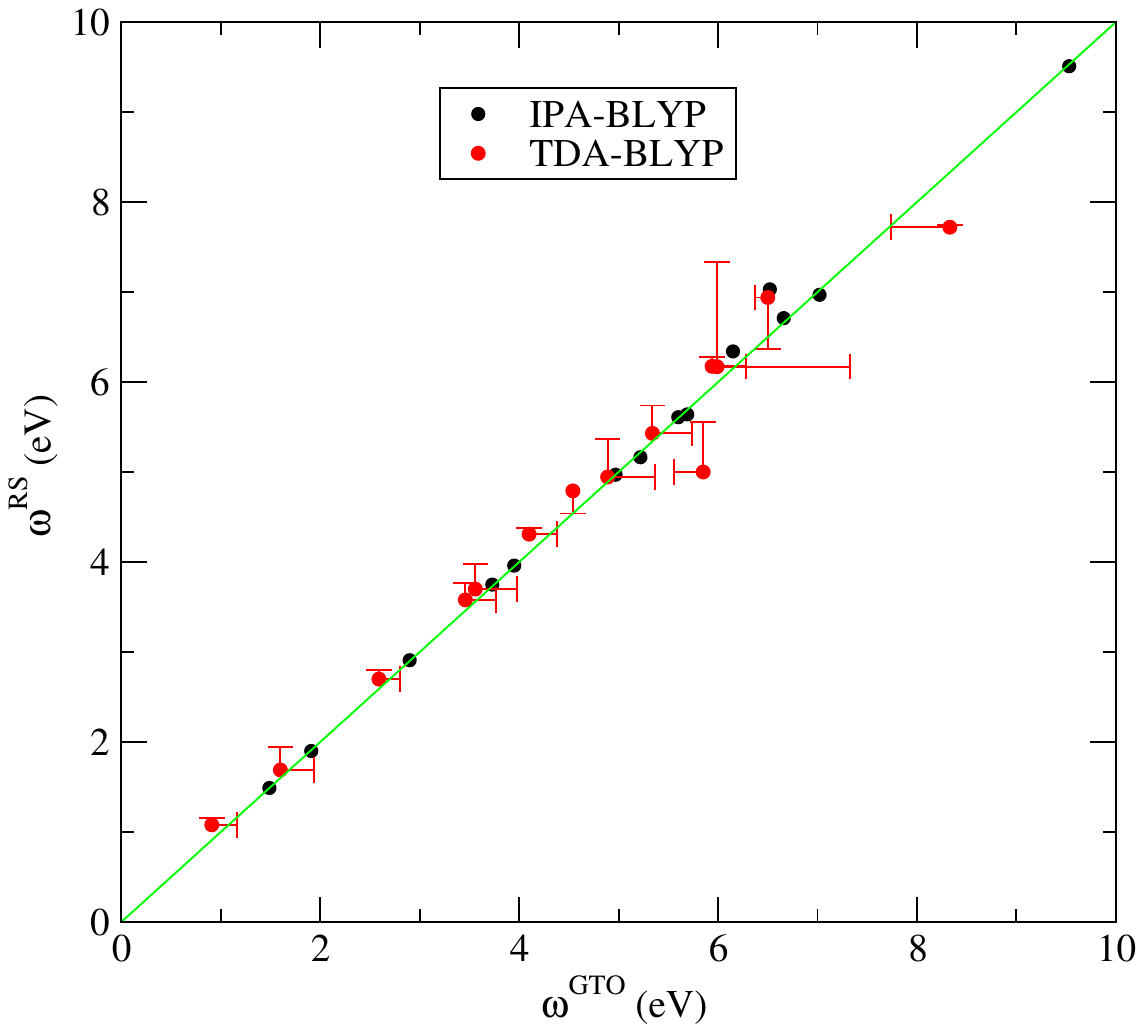}};
    \node at (0.45\textwidth,0.21\textwidth) {b) Triplets};

    \node at (0.55\textwidth,-1.7) {\includegraphics[width=0.18\textwidth]{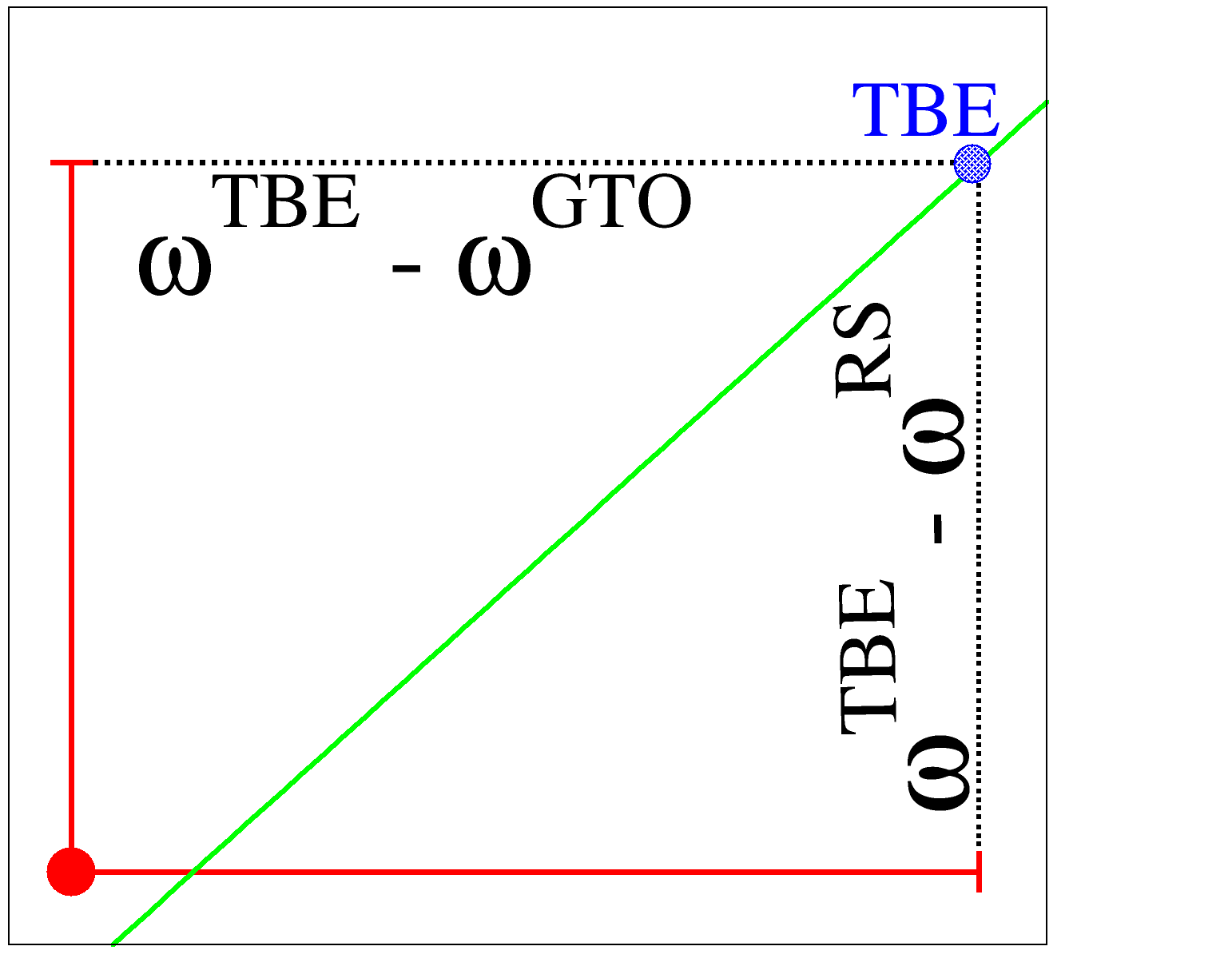}};
\end{tikzpicture}
\caption{Comparison of BLYP excitation energies, from the independent particle approximation (IPA) and TDDFT in the Tamm-Dancoff approximation (TDA), as calculated with GTOs \cite{gould2022single} and real space (RS) for the set of 14 small molecules. The green line represents the line $\omega^{\rm RS} = \omega^{\rm GTO}$, and error bars represent the signed deviation from the ``Theoretical Best Estimate'' (TBE) \cite{loos2018mountaineering}.}
\label{BLYP_Bmark}
\end{figure}

\begin{figure}
  \centering
\begin{tikzpicture}
    \node at (-0.2,0) {\includegraphics[width=.7\textwidth]{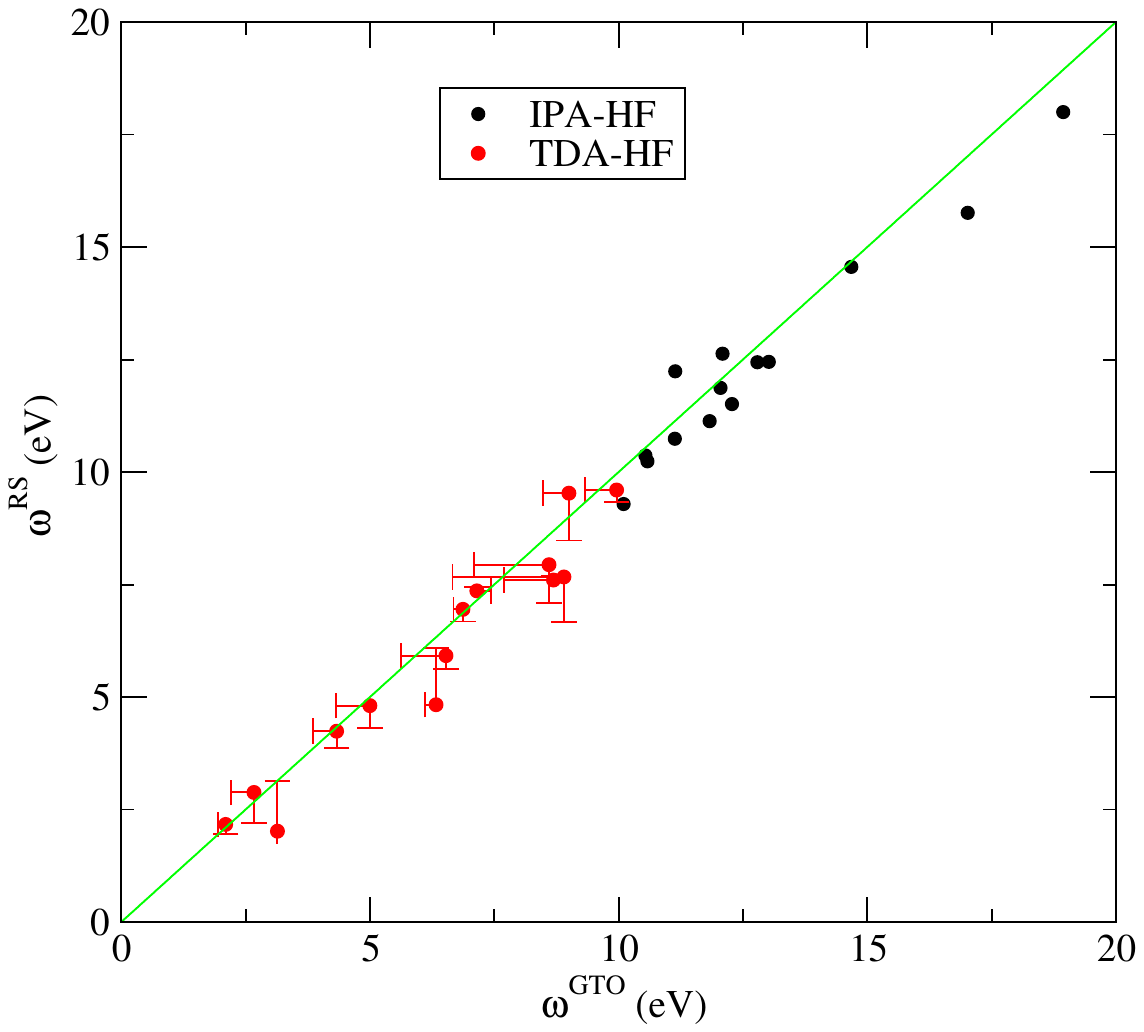}};
    \node at (-0.3,0.21\textwidth) {a) Singlets};

    \node at (0.5\textwidth,0) {\includegraphics[width=0.7\textwidth]{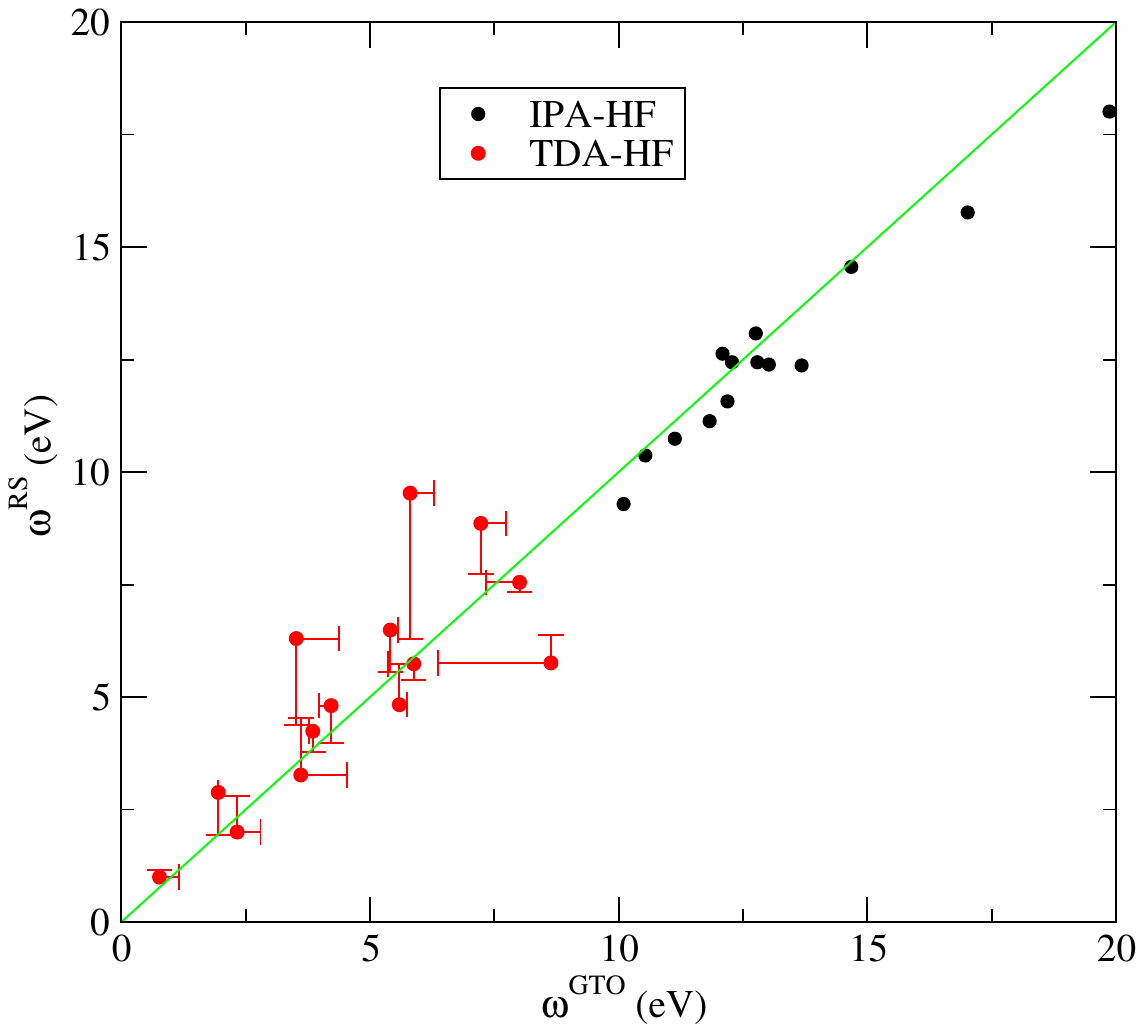}};
    \node at (0.45\textwidth,0.21\textwidth) {b) Triplets};

    \node at (0.55\textwidth,-1.7) {\includegraphics[width=0.18
    \textwidth]{inset.png}};

\end{tikzpicture}
\caption{Comparison of TDHF excitation energies, from the independent particle approximation (IPA) and the Tamm-Dancoff approximation (TDA), as calculated with GTOs \cite{gould2022single} and real space (RS) for the set of 14 small molecules. The green line represents the line $\omega^{\rm RS} = \omega^{\rm GTO}$, and error bars represent the signed deviation from the ``Theoretical Best Estimate'' (TBE) \cite{loos2018mountaineering}.}
\label{HF_Bmark}
\end{figure}

The results are summarized in figures \ref{BLYP_Bmark} and \ref{HF_Bmark}, where the error bars represent the signed deviation from the TBE.
We observe that in the case of triplet excitations, TDHF with GTOs is closer to the TBE than TDHF with real space, whereas for singlet excitations, TDHF with real space is closer to the TBE than TDHF with GTOs. These differing results for TDHF stem from differences in the accuracy of two integrals in equation \ref{RHF_matrix}, and cancellation of errors due to basis set and level of theory. By contrast, TDDFT results show similar deviations from the TBE for both GTO and real space.

\begin{figure}[hbt!]
    \centering
    \includegraphics[width=.7\textwidth]{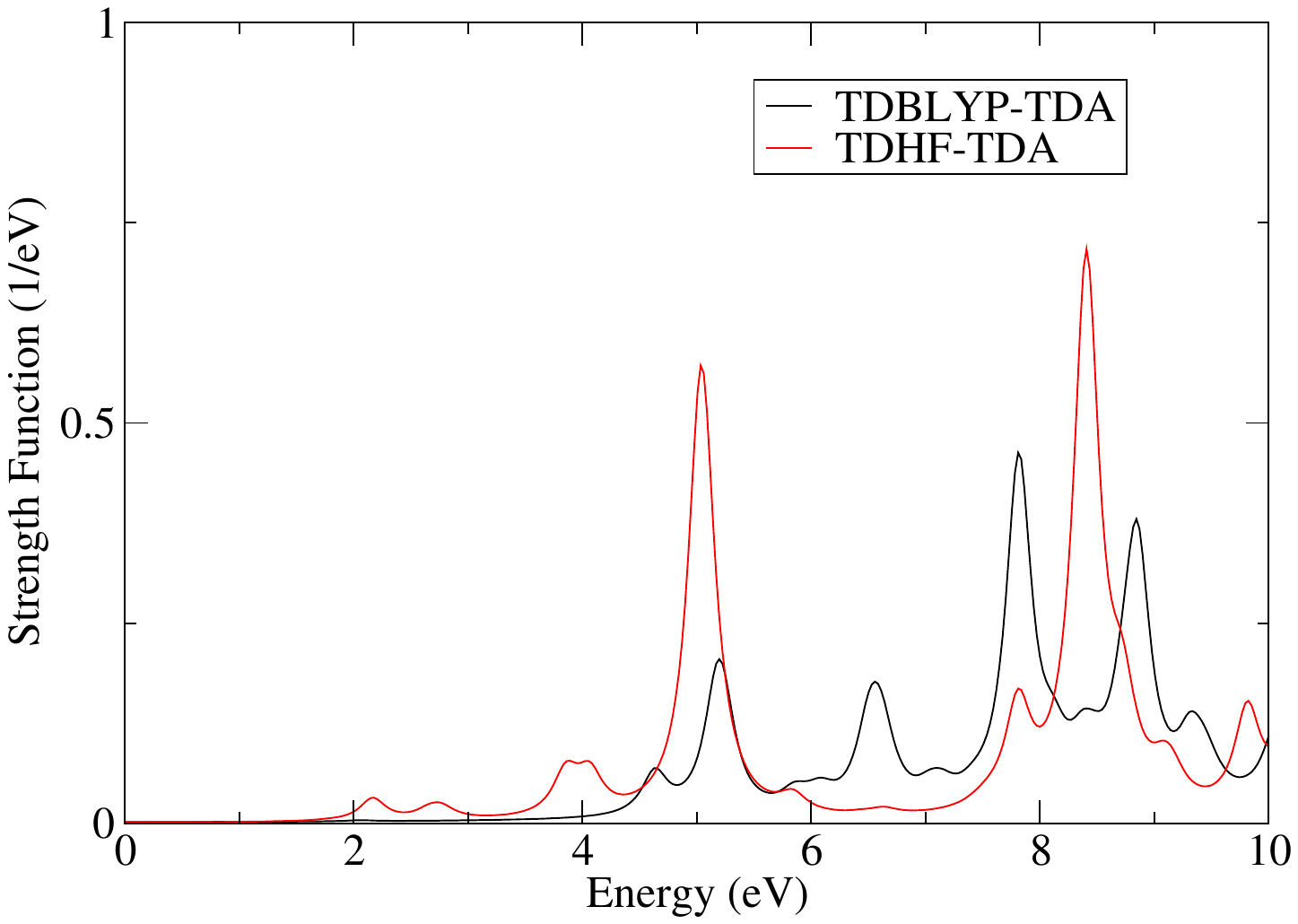}
    \caption{Comparison of absorption spectra of nitrosomethane molecule from TDDFT using BLYP, and TDHF, both using the TDA and $R=10$ \r{A}, $h=0.10$ \r{A}. Lorentzian of width $0.0136$ eV used for broadening.}
    \label{BLYPvsHF_nitroso}
\end{figure}

\begin{figure}[hbt!]
    \centering
    \includegraphics[width=0.7\textwidth]{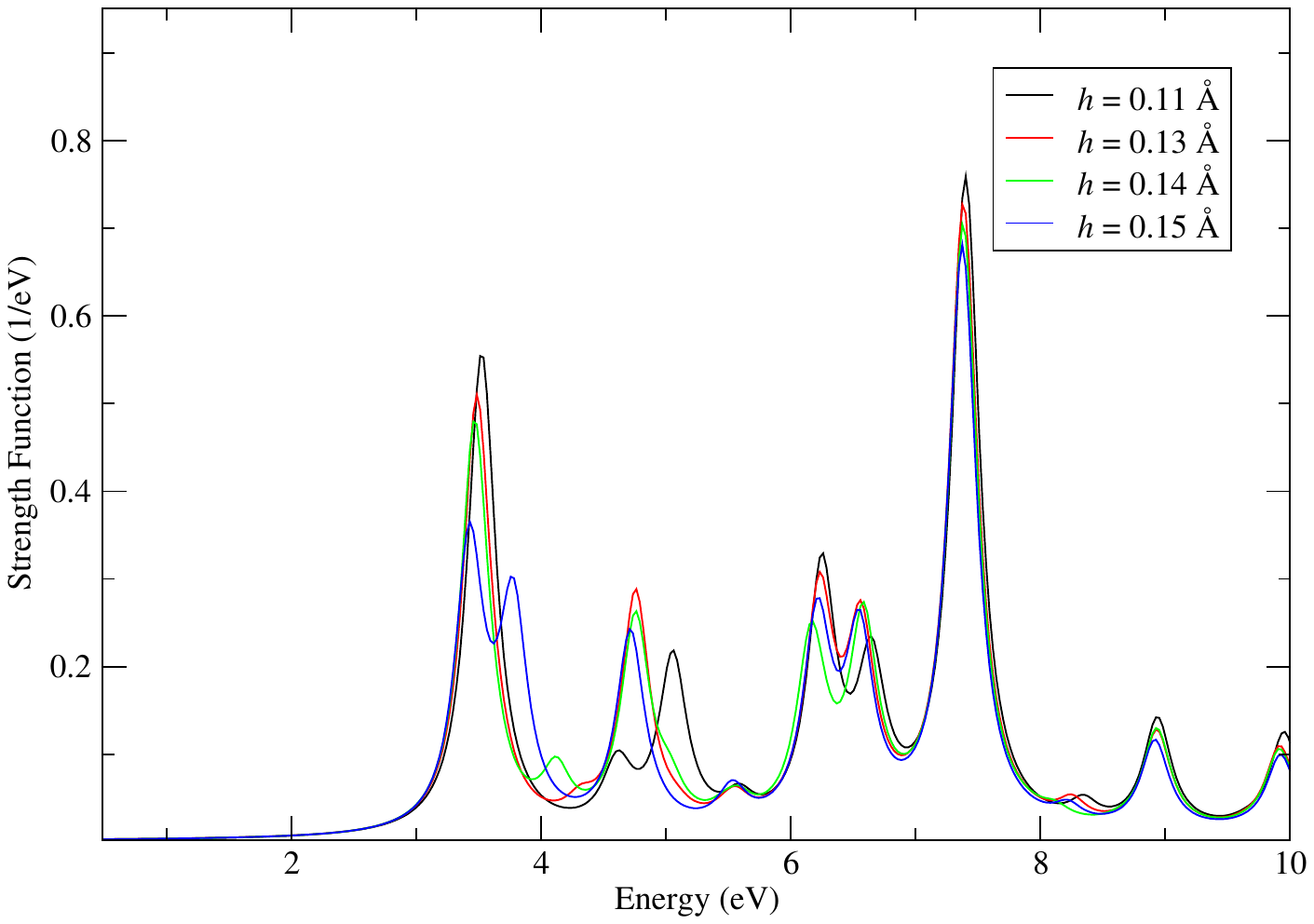}
    \caption{Convergence of absorption spectra of thioformaldehyde molecule with respect to grid spacing from TDHF using TDA, with sphere $R = 5$ \r{A}. Lorentzian of width $0.0136$ eV used for broadening.}
    \label{conv_spec_thio}
\end{figure}

\begin{figure}[H]
     \centering
     \includegraphics[width=.7\textwidth]{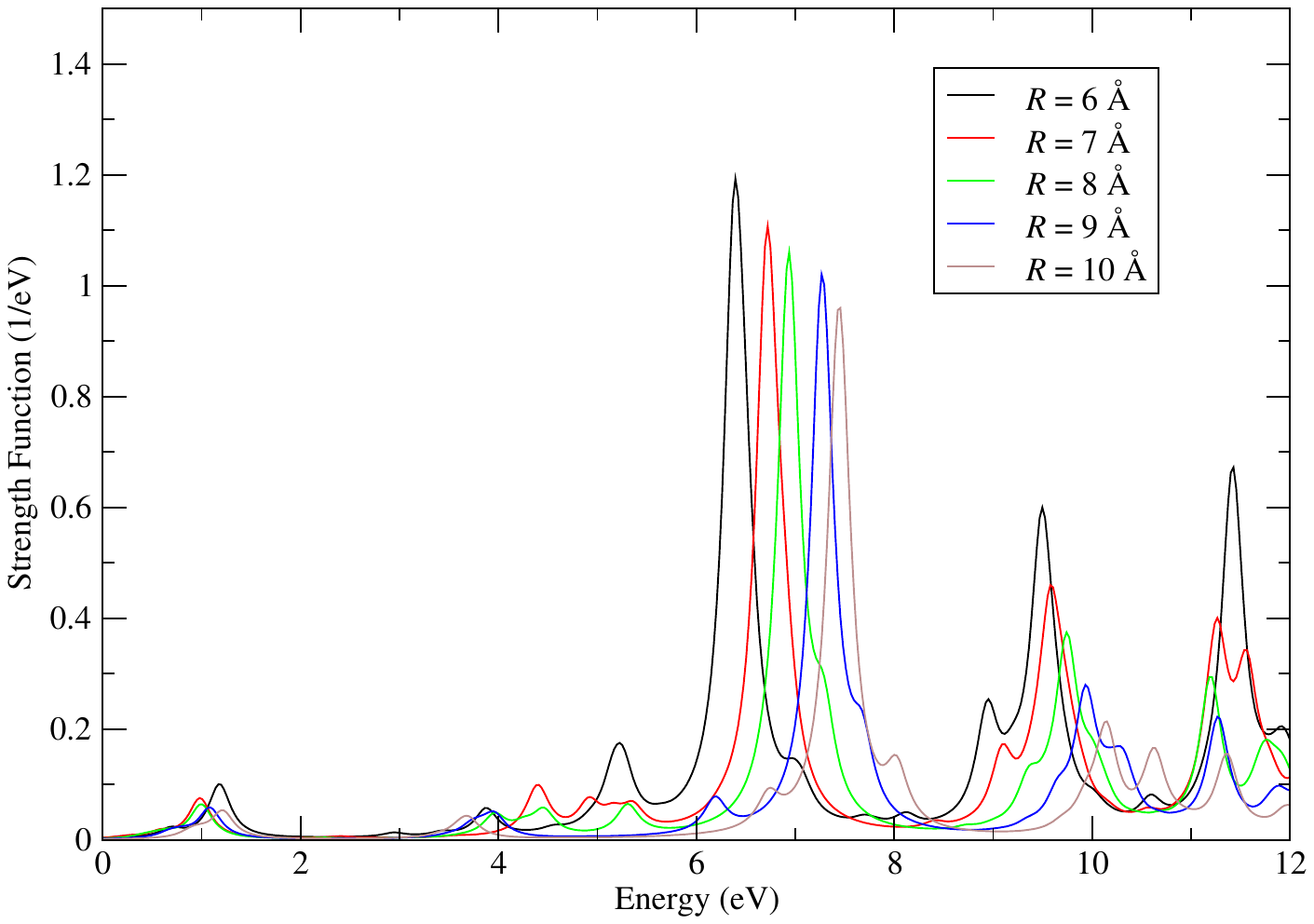}
     \caption{Convergence of absorption spectra of diazomethane molecule with respect to the radius of the simulation sphere obtained from TDHF using TDA, with grid spacing $h=0.10$ \r{A}. Lorentzian of width $0.0136$ eV used for broadening.}
     \label{conv_spec_diazo_R}
\end{figure}

We have also calculated the absorption spectrum using TDHF and compared it to the spectrum from TDDFT. The absorption spectra of the nitrosomethane molecule is shown in figure \ref{BLYPvsHF_nitroso}. In addition, to show that these spectra can be successfully and systematically converged, we have shown the absorption spectrum for thioformaldehyde at different grid spacings in figure \ref{conv_spec_thio}. We see that convergence of occupied eigenvalues to $\sim0.001$ eV and unoccupied eigenvalues to $\sim0.01$ eV (figure \ref{convEval}) leads to convergence of singlet excitation energies within $\sim0.1$ eV: the integrals in equation \ref{RHF_matrix} are more demanding to converge than the eigenvalues. We also demonstrate convergence with respect to the simulation sphere $R$ in figure \ref{conv_spec_diazo_R}, for the absorption spectrum of diazomethane. The low-lying excitation around 1 eV is fairly well converged by $R = 10$\ \AA, but the higher excitations around 7 eV and above need a yet larger sphere because of their Rydberg nature and interaction with the continuum states of the vacuum.

\begin{table}[]
\centering
\begin{tabular}{|l||crcr||crcr|}
\hline
\multirow{3}{*}{} &
  \multicolumn{4}{c||}{Singlets} &
  \multicolumn{4}{c|}{Triplets} \\ \cline{2-9} 
 &
  \multicolumn{2}{c|}{BLYP} &
  \multicolumn{2}{c||}{HF} &
  \multicolumn{2}{c|}{BLYP} &
  \multicolumn{2}{c|}{HF} \\ \cline{2-9} 
 &
  \multicolumn{1}{c|}{IPA} &
  \multicolumn{1}{c|}{TDA} &
  \multicolumn{1}{c|}{IPA} &
  \multicolumn{1}{c||}{TDA} &
  \multicolumn{1}{c|}{IPA} &
  \multicolumn{1}{c|}{TDA} &
  \multicolumn{1}{c|}{IPA} &
  \multicolumn{1}{c|}{TDA} \\ \hline \hline
ME &
  \multicolumn{1}{r|}{-0.02} &
  \multicolumn{1}{r|}{0.06} &
  \multicolumn{1}{r|}{-0.20} &
  -0.12 &
  \multicolumn{1}{r|}{0.04} &
  \multicolumn{1}{r|}{0.05} &
  \multicolumn{1}{r|}{-0.51} &
  0.46 \\ \hline
MAE &
  \multicolumn{1}{r|}{0.04} &
  \multicolumn{1}{r|}{0.07} &
  \multicolumn{1}{r|}{0.36} &
  0.37 &
  \multicolumn{1}{r|}{0.07} &
  \multicolumn{1}{r|}{0.25} &
  \multicolumn{1}{r|}{0.66} &
  1.16 \\ \hline
\end{tabular}
\caption{Mean Error (ME) and Mean Absolute Error (MAE) in eV between real-space (RS) Octopus and Gaussian-type orbital (GTO) Psi4 calculations, for excitation energies from TDDFT (BLYP) and TDHF, in the IPA and TDA approaches.}
\label{RSvsGTOs_Errors}
\end{table}

\begin{table}[]
\footnotesize
\begin{tabular}{|lcr||rrrr||rrrr|}
\hline
\multicolumn{3}{|l||}{Triplets}                                                                                                                                     & \multicolumn{4}{l||}{BLYP}                                                                                                                                                                                                                                                                         & \multicolumn{4}{l|}{HF}                                                                                                                                                                                                                                                                           \\ \hline
\multicolumn{1}{|l|}{Molecule}                                                                      & \multicolumn{1}{l|}{Symmetry}     & \multicolumn{1}{l||}{TBE} & \multicolumn{1}{l|}{\begin{tabular}[c]{@{}l@{}}IPA \\ GTO\end{tabular}} & \multicolumn{1}{l|}{\begin{tabular}[c]{@{}l@{}}IPA\\ RS\end{tabular}} & \multicolumn{1}{l|}{\begin{tabular}[c]{@{}l@{}}TDA \\ GTO\end{tabular}} & \multicolumn{1}{l||}{\begin{tabular}[c]{@{}l@{}}TDA\\ RS\end{tabular}} & \multicolumn{1}{l|}{\begin{tabular}[c]{@{}l@{}}IPA \\ GTO\end{tabular}} & \multicolumn{1}{l|}{\begin{tabular}[c]{@{}l@{}}IPA\\ RS\end{tabular}} & \multicolumn{1}{l|}{\begin{tabular}[c]{@{}l@{}}TDA \\ GTO\end{tabular}} & \multicolumn{1}{l|}{\begin{tabular}[c]{@{}l@{}}TDA\\ RS\end{tabular}} \\ \hline\hline
\multicolumn{1}{|l|}{acetaldehyde}                                                                  & \multicolumn{1}{c|}{A$''$}        & 3.98                     & \multicolumn{1}{r|}{3.95}                                               & \multicolumn{1}{r|}{3.96}                                             & \multicolumn{1}{r|}{3.56}                                               & 3.70                                                                  & \multicolumn{1}{r|}{12.79}                                              & \multicolumn{1}{r|}{12.44}                                            & \multicolumn{1}{r|}{4.22}                                               & 4.81                                                                  \\ \hline
\multicolumn{1}{|l|}{acetylene}                                                                     & \multicolumn{1}{c|}{A$_{\rm u}$}  & 5.56                     & \multicolumn{1}{r|}{6.66}                                               & \multicolumn{1}{r|}{6.71}                                             & \multicolumn{1}{r|}{5.85}                                               & 5.00                                                                  & \multicolumn{1}{r|}{13.68}                                              & \multicolumn{1}{r|}{12.37}                                            & \multicolumn{1}{r|}{5.41}                                               & 6.49                                                                  \\ \hline
\multicolumn{1}{|l|}{ammonia}                                                                       & \multicolumn{1}{c|}{A$''$}        & 6.37                     & \multicolumn{1}{r|}{6.52}                                               & \multicolumn{1}{r|}{7.03}                                             & \multicolumn{1}{r|}{6.50}                                               & 6.94                                                                  & \multicolumn{1}{r|}{13.02}                                              & \multicolumn{1}{r|}{12.39}                                            & \multicolumn{1}{r|}{8.64}                                               & 5.76                                                                  \\ \hline
\multicolumn{1}{|l|}{carbon monoxide}                                                              & \multicolumn{1}{c|}{B$_1$}        & 6.28                     & \multicolumn{1}{r|}{7.02}                                               & \multicolumn{1}{r|}{6.97}                                             & \multicolumn{1}{r|}{5.94}                                               & 6.18                                                                  & \multicolumn{1}{r|}{17.02}                                              & \multicolumn{1}{r|}{15.77}                                            & \multicolumn{1}{r|}{5.81}                                               & 9.53                                                                  \\ \hline
\multicolumn{1}{|l|}{cyclopropene}                                                                  & \multicolumn{1}{c|}{B$_2$}        & 4.38                     & \multicolumn{1}{r|}{4.97}                                               & \multicolumn{1}{r|}{4.97}                                             & \multicolumn{1}{r|}{4.10}                                               & 4.31                                                                  & \multicolumn{1}{r|}{12.19}                                              & \multicolumn{1}{r|}{11.57}                                            & \multicolumn{1}{r|}{3.52}                                               & 6.30                                                                  \\ \hline
\multicolumn{1}{|l|}{diazomethane}                                                                  & \multicolumn{1}{c|}{A$_2$}        & 2.80                     & \multicolumn{1}{r|}{2.90}                                               & \multicolumn{1}{r|}{2.91}                                             & \multicolumn{1}{r|}{2.59}                                               & 2.70                                                                  & \multicolumn{1}{r|}{10.10}                                              & \multicolumn{1}{r|}{9.29}                                             & \multicolumn{1}{r|}{2.33}                                               & 2.00                                                                  \\ \hline
\multicolumn{1}{|l|}{dinitrogen}                                                                    & \multicolumn{1}{c|}{A$_{\rm u}$}  & 7.74                     & \multicolumn{1}{r|}{9.53}                                               & \multicolumn{1}{r|}{9.51}                                             & \multicolumn{1}{r|}{8.33}                                               & 7.72                                                                  & \multicolumn{1}{r|}{19.87}                                              & \multicolumn{1}{r|}{18.01}                                            & \multicolumn{1}{r|}{7.23}                                               & 8.86                                                                  \\ \hline
\multicolumn{1}{|l|}{ethylene}                                                                      & \multicolumn{1}{c|}{B$_{\rm 2u}$} & 4.54                     & \multicolumn{1}{r|}{5.60}                                               & \multicolumn{1}{r|}{5.61}                                             & \multicolumn{1}{r|}{4.54}                                               & 4.79                                                                  & \multicolumn{1}{r|}{12.76}                                              & \multicolumn{1}{r|}{13.08}                                            & \multicolumn{1}{r|}{3.61}                                               & 3.27                                                                  \\ \hline
\multicolumn{1}{|l|}{formamide}                                                                     & \multicolumn{1}{c|}{A$''$}        & 5.37                     & \multicolumn{1}{r|}{5.22}                                               & \multicolumn{1}{r|}{5.16}                                             & \multicolumn{1}{r|}{4.89}                                               & 4.95                                                                  & \multicolumn{1}{r|}{12.09}                                              & \multicolumn{1}{r|}{12.63}                                            & \multicolumn{1}{r|}{5.88}                                               & 5.74                                                                  \\ \hline
\multicolumn{1}{|l|}{hydrogen sulphide}                                                             & \multicolumn{1}{c|}{A$_2$}        & 5.74                     & \multicolumn{1}{r|}{5.69}                                               & \multicolumn{1}{r|}{5.64}                                             & \multicolumn{1}{r|}{5.34}                                               & 5.43                                                                  & \multicolumn{1}{r|}{11.83}                                              & \multicolumn{1}{r|}{11.13}                                            & \multicolumn{1}{r|}{5.59}                                               & 4.83                                                                  \\ \hline
\multicolumn{1}{|l|}{ketene}                                                                        & \multicolumn{1}{c|}{A$_2$}        & 3.77                     & \multicolumn{1}{r|}{3.73}                                               & \multicolumn{1}{r|}{3.75}                                             & \multicolumn{1}{r|}{3.46}                                               & 3.58                                                                  & \multicolumn{1}{r|}{11.13}                                              & \multicolumn{1}{r|}{10.74}                                            & \multicolumn{1}{r|}{3.85}                                               & 4.24                                                                  \\ \hline
\multicolumn{1}{|l|}{nitrosomethane}                                                                & \multicolumn{1}{c|}{A$''$}        & 1.16                     & \multicolumn{1}{r|}{1.49}                                               & \multicolumn{1}{r|}{1.49}                                             & \multicolumn{1}{r|}{0.91}                                               & 1.08                                                                  & \multicolumn{1}{r|}{12.28}                                              & \multicolumn{1}{r|}{12.44}                                            & \multicolumn{1}{r|}{0.77}                                               & 1.00                                                                  \\ \hline
\multicolumn{1}{|l|}{thioformaldehyde}                                                              & \multicolumn{1}{c|}{A$_2$}        & 1.94                     & \multicolumn{1}{r|}{1.91}                                               & \multicolumn{1}{r|}{1.90}                                             & \multicolumn{1}{r|}{1.60}                                               & 1.69                                                                  & \multicolumn{1}{r|}{10.54}                                              & \multicolumn{1}{r|}{10.37}                                            & \multicolumn{1}{r|}{1.95}                                               & 2.88                                                                  \\ \hline
\multicolumn{1}{|l|}{water}                                                                         & \multicolumn{1}{c|}{B$_1$}        & 7.33                     & \multicolumn{1}{r|}{6.15}                                               & \multicolumn{1}{r|}{6.34}                                             & \multicolumn{1}{r|}{5.99}                                               & 6.17                                                                  & \multicolumn{1}{r|}{14.68}                                              & \multicolumn{1}{r|}{14.56}                                            & \multicolumn{1}{r|}{8.01}                                               & 7.55                                                                  \\ \hline\hline
\multicolumn{1}{|c|}{\multirow{2}{*}{\begin{tabular}[c]{@{}c@{}}Deviation from\\ GTO\end{tabular}}} & \multicolumn{2}{c||}{ME}                                      & \multicolumn{1}{r|}{\multirow{2}{*}{}}                                  & \multicolumn{1}{r|}{0.04}                                             & \multicolumn{1}{r|}{\multirow{2}{*}{}}                                  & 0.05                                                                  & \multicolumn{1}{r|}{\multirow{2}{*}{}}                                  & \multicolumn{1}{r|}{-0.51}                                            & \multicolumn{1}{r|}{\multirow{2}{*}{}}                                  & 0.46                                                                  \\ \cline{2-3} \cline{5-5} \cline{7-7} \cline{9-9} \cline{11-11} 
\multicolumn{1}{|c|}{}                                                                              & \multicolumn{2}{c||}{MAE}                                     & \multicolumn{1}{r|}{}                                                   & \multicolumn{1}{r|}{0.07}                                             & \multicolumn{1}{r|}{}                                                   & 0.25                                                                  & \multicolumn{1}{r|}{}                                                   & \multicolumn{1}{r|}{0.66}                                             & \multicolumn{1}{r|}{}                                                   & 1.16                                                                  \\ \hline \hline
\multicolumn{1}{|c|}{\multirow{2}{*}{\begin{tabular}[c]{@{}c@{}}Deviation from\\ TBE\end{tabular}}} & \multicolumn{2}{c||}{ME}                                      & \multicolumn{2}{r|}{\multirow{2}{*}{}}                                                                                                          & \multicolumn{1}{r|}{-0.24}                                              & -0.20                                                                 & \multicolumn{2}{r|}{\multirow{2}{*}{}}                                                                                                          & \multicolumn{1}{r|}{-0.01}                                              & 0.45                                                                  \\ \cline{2-3} \cline{6-7} \cline{10-11} 
\multicolumn{1}{|c|}{}                                                                              & \multicolumn{2}{c||}{MAE}                                     & \multicolumn{2}{r|}{}                                                                                                                           & \multicolumn{1}{r|}{0.38}                                               & 0.31                                                                  & \multicolumn{2}{r|}{}                                                                                                                           & \multicolumn{1}{r|}{0.55}                                               & 0.99                                                                  \\ \hline
\end{tabular}
\caption{Triplet excitation energies (eV) from TDDFT and TDHF of various molecules, with the irreducible representations given for each excitation. Real-space (RS) results are compared to a GTO dataset  \cite{gould2022single} for each method, and also to the higher-level theory TBE dataset  \cite{loos2018mountaineering}.}
\label{Bmark_trips}
\end{table}

\begin{table}[]
\footnotesize
\begin{tabular}{|lcr||rrrr||rrrr|}
\hline
\multicolumn{3}{|l||}{Singlets}                                                                                                                                    & \multicolumn{4}{l||}{BLYP}                                                                                                                                                                                                                                                                         & \multicolumn{4}{l|}{HF}                                                                                                                                                                                                                                                                           \\ \hline\hline
\multicolumn{1}{|l|}{Molecule}                                                                      & \multicolumn{1}{l|}{Symmetry}    & \multicolumn{1}{l||}{TBE} & \multicolumn{1}{l|}{\begin{tabular}[c]{@{}l@{}}IPA \\ GTO\end{tabular}} & \multicolumn{1}{l|}{\begin{tabular}[c]{@{}l@{}}IPA\\ RS\end{tabular}} & \multicolumn{1}{l|}{\begin{tabular}[c]{@{}l@{}}TDA \\ GTO\end{tabular}} & \multicolumn{1}{l||}{\begin{tabular}[c]{@{}l@{}}TDA\\ RS\end{tabular}} & \multicolumn{1}{l|}{\begin{tabular}[c]{@{}l@{}}IPA \\ GTO\end{tabular}} & \multicolumn{1}{l|}{\begin{tabular}[c]{@{}l@{}}IPA\\ RS\end{tabular}} & \multicolumn{1}{l|}{\begin{tabular}[c]{@{}l@{}}TDA \\ GTO\end{tabular}} & \multicolumn{1}{l|}{\begin{tabular}[c]{@{}l@{}}TDA\\ RS\end{tabular}} \\ \hline
\multicolumn{1}{|l|}{acetaldehyde}                                                                  & \multicolumn{1}{c|}{A$''$}       & 4.31                     & \multicolumn{1}{r|}{3.95}                                               & \multicolumn{1}{r|}{3.93}                                             & \multicolumn{1}{r|}{4.17}                                               & 4.25                                                                  & \multicolumn{1}{r|}{12.79}                                              & \multicolumn{1}{r|}{12.44}                                            & \multicolumn{1}{r|}{5.00}                                               & 4.81                                                                  \\ \hline
\multicolumn{1}{|l|}{acetylene}                                                                     & \multicolumn{1}{c|}{B$_{\rm2g}$} & 7.10                     & \multicolumn{1}{r|}{7.01}                                               & \multicolumn{1}{r|}{6.65}                                             & \multicolumn{1}{r|}{7.08}                                               & 6.65                                                                  & \multicolumn{1}{r|}{12.05}                                              & \multicolumn{1}{r|}{11.87}                                            & \multicolumn{1}{r|}{8.60}                                               & 7.94                                                                  \\ \hline
\multicolumn{1}{|l|}{ammonia}                                                                       & \multicolumn{1}{c|}{A$''$}       & 6.66                     & \multicolumn{1}{r|}{6.52}                                               & \multicolumn{1}{r|}{6.68}                                             & \multicolumn{1}{r|}{6.53}                                               & 6.68                                                                  & \multicolumn{1}{r|}{13.02}                                              & \multicolumn{1}{r|}{12.45}                                            & \multicolumn{1}{r|}{8.90}                                               & 7.67                                                                  \\ \hline
\multicolumn{1}{|l|}{carbon monoxide}                                                              & \multicolumn{1}{c|}{B$_1$}       & 8.48                     & \multicolumn{1}{r|}{7.02}                                               & \multicolumn{1}{r|}{7.01}                                             & \multicolumn{1}{r|}{8.39}                                               & 8.59                                                                  & \multicolumn{1}{r|}{17.02}                                              & \multicolumn{1}{r|}{15.76}                                            & \multicolumn{1}{r|}{9.00}                                               & 9.53                                                                  \\ \hline
\multicolumn{1}{|l|}{cyclopropene}                                                                  & \multicolumn{1}{c|}{B$_1$}       & 6.68                     & \multicolumn{1}{r|}{5.70}                                               & \multicolumn{1}{r|}{5.63}                                             & \multicolumn{1}{r|}{5.70}                                               & 5.59                                                                  & \multicolumn{1}{r|}{10.58}                                              & \multicolumn{1}{r|}{10.24}                                            & \multicolumn{1}{r|}{6.87}                                               & 6.95                                                                  \\ \hline
\multicolumn{1}{|l|}{diazomethane}                                                                  & \multicolumn{1}{c|}{A$_2$}       & 3.13                     & \multicolumn{1}{r|}{2.90}                                               & \multicolumn{1}{r|}{2.89}                                             & \multicolumn{1}{r|}{3.00}                                               & 3.07                                                                  & \multicolumn{1}{r|}{10.10}                                              & \multicolumn{1}{r|}{9.29}                                             & \multicolumn{1}{r|}{3.14}                                               & 2.02                                                                  \\ \hline
\multicolumn{1}{|l|}{dinitrogen}                                                                    & \multicolumn{1}{c|}{B$_{\rm2g}$} & 9.33                     & \multicolumn{1}{r|}{8.31}                                               & \multicolumn{1}{r|}{8.30}                                             & \multicolumn{1}{r|}{9.16}                                               & 9.38                                                                  & \multicolumn{1}{r|}{18.94}                                              & \multicolumn{1}{r|}{18.00}                                            & \multicolumn{1}{r|}{9.96}                                               & 9.60                                                                  \\ \hline
\multicolumn{1}{|l|}{ethylene}                                                                      & \multicolumn{1}{c|}{B$_{\rm3u}$} & 7.44                     & \multicolumn{1}{r|}{6.23}                                               & \multicolumn{1}{r|}{6.16}                                             & \multicolumn{1}{r|}{6.24}                                               & 6.14                                                                  & \multicolumn{1}{r|}{11.14}                                              & \multicolumn{1}{r|}{12.24}                                            & \multicolumn{1}{r|}{7.15}                                               & 7.36                                                                  \\ \hline
\multicolumn{1}{|l|}{formamide}                                                                     & \multicolumn{1}{c|}{A$''$}       & 5.63                     & \multicolumn{1}{r|}{5.22}                                               & \multicolumn{1}{r|}{5.19}                                             & \multicolumn{1}{r|}{5.39}                                               & 5.44                                                                  & \multicolumn{1}{r|}{12.09}                                              & \multicolumn{1}{r|}{12.63}                                            & \multicolumn{1}{r|}{6.53}                                               & 5.92                                                                  \\ \hline
\multicolumn{1}{|l|}{hydrogen sulphide}                                                             & \multicolumn{1}{c|}{A$_2$}       & 6.10                     & \multicolumn{1}{r|}{5.68}                                               & \multicolumn{1}{r|}{5.55}                                             & \multicolumn{1}{r|}{5.80}                                               & 5.74                                                                  & \multicolumn{1}{r|}{11.83}                                              & \multicolumn{1}{r|}{11.13}                                            & \multicolumn{1}{r|}{6.33}                                               & 4.83                                                                  \\ \hline
\multicolumn{1}{|l|}{ketene}                                                                        & \multicolumn{1}{c|}{A$_2$}       & 3.86                     & \multicolumn{1}{r|}{3.72}                                               & \multicolumn{1}{r|}{3.73}                                             & \multicolumn{1}{r|}{3.81}                                               & 3.88                                                                  & \multicolumn{1}{r|}{11.13}                                              & \multicolumn{1}{r|}{10.74}                                            & \multicolumn{1}{r|}{4.33}                                               & 4.24                                                                  \\ \hline
\multicolumn{1}{|l|}{nitrosomethane}                                                                & \multicolumn{1}{c|}{A$''$}       & 1.95                     & \multicolumn{1}{r|}{1.49}                                               & \multicolumn{1}{r|}{1.45}                                             & \multicolumn{1}{r|}{1.93}                                               & 2.05                                                                  & \multicolumn{1}{r|}{12.28}                                              & \multicolumn{1}{r|}{11.51}                                            & \multicolumn{1}{r|}{2.10}                                               & 2.17                                                                  \\ \hline
\multicolumn{1}{|l|}{thioformaldehyde}                                                              & \multicolumn{1}{c|}{A$_2$}       & 2.20                     & \multicolumn{1}{r|}{1.91}                                               & \multicolumn{1}{r|}{1.88}                                             & \multicolumn{1}{r|}{2.15}                                               & 2.20                                                                  & \multicolumn{1}{r|}{10.54}                                              & \multicolumn{1}{r|}{10.37}                                            & \multicolumn{1}{r|}{2.67}                                               & 2.88                                                                  \\ \hline
\multicolumn{1}{|l|}{water}                                                                         & \multicolumn{1}{c|}{B$_1$}       & 7.70                     & \multicolumn{1}{r|}{6.15}                                               & \multicolumn{1}{r|}{6.08}                                             & \multicolumn{1}{r|}{6.23}                                               & 6.19                                                                  & \multicolumn{1}{r|}{14.68}                                              & \multicolumn{1}{r|}{14.56}                                            & \multicolumn{1}{r|}{8.69}                                               & 7.60                                                                  \\ \hline\hline
\multicolumn{1}{|c|}{\multirow{2}{*}{\begin{tabular}[c]{@{}c@{}}Deviation from\\ GTO\end{tabular}}} & \multicolumn{2}{c||}{ME}                                     & \multicolumn{1}{r|}{\multirow{2}{*}{}}                                  & \multicolumn{1}{r|}{-0.02}                                            & \multicolumn{1}{r|}{\multirow{2}{*}{}}                                  & 0.06                                                                  & \multicolumn{1}{r|}{\multirow{2}{*}{}}                                  & \multicolumn{1}{r|}{-0.20}                                            & \multicolumn{1}{r|}{\multirow{2}{*}{}}                                  & -0.12                                                                 \\ \cline{2-3} \cline{5-5} \cline{7-7} \cline{9-9} \cline{11-11} 
\multicolumn{1}{|c|}{}                                                                              & \multicolumn{2}{c||}{MAE}                                    & \multicolumn{1}{r|}{}                                                   & \multicolumn{1}{r|}{0.04}                                             & \multicolumn{1}{r|}{}                                                   & 0.07                                                                  & \multicolumn{1}{r|}{}                                                   & \multicolumn{1}{r|}{0.36}                                             & \multicolumn{1}{r|}{}                                                   & 0.37                                                                  \\ \hline \hline
\multicolumn{1}{|c|}{\multirow{2}{*}{\begin{tabular}[c]{@{}c@{}}Deviation from\\ TBE\end{tabular}}} & \multicolumn{2}{c||}{ME}                                     & \multicolumn{2}{r|}{\multirow{2}{*}{}}                                                                                                          & \multicolumn{1}{r|}{-0.36}                                              & -0.34                                                                 & \multicolumn{2}{r|}{\multirow{2}{*}{}}                                                                                                          & \multicolumn{1}{r|}{0.62}                                               & 0.21                                                                  \\ \cline{2-3} \cline{6-7} \cline{10-11} 
\multicolumn{1}{|c|}{}                                                                              & \multicolumn{2}{c||}{MAE}                                    & \multicolumn{2}{r|}{}                                                                                                                           & \multicolumn{1}{r|}{0.36}                                               & 0.38                                                                  & \multicolumn{2}{r|}{}                                                                                                                           & \multicolumn{1}{r|}{0.66}                                               & 0.58                                                                  \\ \hline
\end{tabular}
\caption{Singlet excitation energies (eV) from TDDFT and TDHF of various molecules, with irreducible representations given for each excitation. Real-space (RS) results are compared to a GTO dataset  \cite{gould2022single} for each method, and also to the higher-level theory TBE dataset  \cite{loos2018mountaineering}.}
\label{Bmark_sings}
\end{table}


\section{Conclusion}
In this work, we have demonstrated a new implementation of linear-response TDHF in the real-space Octopus code, working in the Tamm-Dancoff approximation. For a benchmark set of small molecules, we find good agreement with results from GTOs. The discrepancy between RS and GTOs for TDHF is significantly larger than for TD-BLYP calculations, showing the more sensitive nature of the TDHF integrals. It is also larger for triplets than for singlets. The discrepancy is largely accounted for by discrepancies in the HF eigenvalues, confirming that our TDHF implementation is comparable to the pre-existing HF implementation in Octopus. These discrepancies between RS and GTOs are attributed to the unavoidable use of pseudopotentials in real space; the fact that these pseudopotentials for TDHF are generated with LDA rather than HF; possible differences in degree of convergence of the RS and GTO calculations; and different treatment of Rydberg excitations. We note that in general Rydberg excitations are better described in real space than in GTOs, where their energies may be overestimated, as seen by the negative mean error of RS vs GTOs in many cases (Table \ref{RSvsGTOs_Errors}). Although we have only shown results from RHF here, we have also implemented TDHF for UHF and confirmed that the set of excitations obtained from RHF and UHF is identical for these closed-shell molecules. Open-shell systems require a more sophisticated handling of spin states \cite{Casanova2020} and is beyond the scope of this study. We find that convergence with respect to the real-space domain's radius and spacing is harder than for DFT, but still achievable, for both excitation energies and absorption spectra.

Since a linear-response excited-state calculation requires an accurate ground-state calculation first, we examined how to get the HF ground state in real space in a practical amount of time. We find that starting with a solution from DFT-LDA as a initial guess instead of LCAO directly from HF speeds up the ground-state convergence considerably. The convergence might not be possible at all in some cases without this. We also find that the best way to achieve fast convergence is by mixing states. Linear mixing of potential or density also provides a considerable increase in the rate of convergence compared to the Broyden scheme, which doesn't seem to facilitate the convergence of the HF ground state.

The calculation of HF ground state also requires the calculation of exact exchange which is particularly time-consuming. We have analyzed the use of ACE, an approach to speeding up the exact-exchange calculation which can also be used for hybrid DFT. Using the existing implementation of ACE in Octopus, we found that for our benchmark set of molecules, the use of ACE sped up the time per SCF iteration by a factor of 7-14. We have also confirmed that the results using ACE match the results without it.

The parallel performance of our TDHF implementation relies on the parallel performance of the Poisson solver \cite{garcia2014survey} for exchange matrix elements, using Octopus' efficient parallelization over real-space domains \cite{andrade2012time}. TDHF can be run also with a level of parallelization over response matrix elements (equations 4, 5, 7) like in the Casida approach to TDDFT \cite{andrade2015real}.

This work adds TDHF to the list of features of the Octopus code, allowing users to compare various approaches in the same platform. This functionality will be available in a forthcoming release. We also note that with some further developments to the code, the ingredients from TDHF can now be used to implement hybrid functionals in TDDFT linear-response calculations \cite{dreuw2005single} and other theories like ensemble DFT \cite{DEC,gould2022single}.

\section*{Acknowledgments}
We acknowledge Nicolas Tancogne-Dejean for the implementation of ACE in Octopus. This work was supported by the U.S. Department of Energy, Office of Science, Basic Energy Sciences, CTC and CPIMS Programs, under Award DE-SC0019053; Cottrell Scholar Award No. 26921, a program of Research Corporation for Science Advancement; and the UC Merced Academic Senate Faculty Research Grants Program. Computational resources were provided by the MERCED (MRI-1429783) and Pinnacles (MRI-2019144) clusters funded by the National Science Foundation, at Cyberinfrastructure and Research Technologies (CIRT), University of California, Merced.

\section*{References}

\bibliographystyle{iopart-num}
\bibliography{refs}

\end{document}